REVIEW

## The quest for a universal density functional: The accuracy of density functionals across a broad spectrum of databases in chemistry and physics

Roberto Peverati and Donald G. Truhlar*

*Department of Chemistry, Chemical Theory Center, and Supercomputing Institute, University of Minnesota, Minneapolis, MN 55455-0431, USA*

Kohn–Sham density functional theory is in principle an exact formulation of quantum mechanical electronic structure theory, but in practice we have to rely on approximate exchange–correlation (xc) functionals. The objective of our work has been to design an xc functional with broad accuracy across as wide an expanse of chemistry and physics as possible, leading—as a long-range goal—to a functional with good accuracy for all problems, i.e., a universal functional. To guide our path toward that goal and to measure our progress, we have developed—building on earlier work in our group—a set of databases of reference data for a variety of energetic and structural properties in chemistry and physics. These databases include energies of molecular processes such as atomization, complexation, proton addition, and ionization; they also include molecular geometries and solid-state lattice constants, chemical reaction barrier heights, and cohesive energies and band gaps of solids. For the present paper we gather many of these databases into four comprehensive databases, two with 384 energetic data for chemistry and solid-state physics and another two with 68 structural data for chemistry and solid-state physics, and we test 2 wave function methods and 77 density functionals (12 Minnesota meta functionals and 65 others) in a consistent way across this same broad set of data. We especially highlight the Minnesota density functionals, but the results have broader implications in that one may see the successes and failures of many kinds of density functionals when they are all applied to the same data. Therefore the results provide a status report on the quest for a universal functional.

*truhlar@umn.edu

## 1. Introduction

Density functional theory (DFT) [1] has enabled electronic structure theory to be applied to materials and complex chemical problems with an accuracy unobtainable by any other approach. This makes DFT useful for modeling, prediction, design, and understanding. Mainstream DFT applications are based on Kohn–Sham theory [2] and its various generalizations, especially the spin-polarized formulation [3] and the generalization to treat a portion of the exchange energy by Hartree–Fock (HF) theory [4]. Kohn-Sham DFT is formally exact, but it involves a functional called the exchange-correlation functional—or, for simplicity, the density functional—and the exact form of this functional is unknown and essentially unknowable. The situation is well summarized by the following quotation:

> "DFT is the method of choice for first principles quantum chemical calculations of the electronic structure and properties of many molecular and solid systems. With the exact exchange-correlation functional, DFT would take into full account all complex many-body effects at a computation cost characteristic of mean field approximations. Unfortunately, the exact exchange-correlation functional is unknown, making it essential to pursue the quest of finding more accurate and reliable functionals."[5]

The present article is concerned with practical approximations to the exact exchange-correlation functional—such density functional approximations are usually just called density functionals, and we will use that terminology here. A considerable amount of research, mainly over the last 30 years, has gone into developing better density functionals, and the best available functional for one application is often not the best for another. So it is of great practical importance to learn which density functionals perform well for which applications and to attempt to design density functionals that are as universally successful as possible. This has been a goal of our recent work, and the present article provides one way to see how much progress we and others have made. In particular, the present article involves a test of 12 Minnesota meta functionals and 65 other exchange-correlation functionals against a common database of 451 data representing molecular energetic and structural data and solid-state energetic and structural data.

## 2. Approximations to the exchange-correlation functional

This section introduces the most common types of approximations used in modern Kohn–Sham DFT calculations. Kohn–Sham theory involves parameterizing the electron density by a Slater determinant, which enforces the Pauli exclusion principle and allows one to calculate the so-called noninteracting kinetic energy from the orbitals of the Slater determinant as if it were a wave function. The Kohn-Sham orbitals satisfy a set of coupled differential equations similar to the HF equations of wave function theory (WFT) but containing the exchange-correlation potential instead of the HF exchange potential. The exchange-correlation potential approximates the exchange and adds electron correlation and the electron kinetic energy beyond the noninteracting part.

Kohn–Sham theory writes the energy as a sum of (i) the noninteracting kinetic energy, (ii) the classical Coulomb energy of the electrons, consisting of their interaction with the nuclei and any external field that may be present (in the rest of this article we assume no such external field is present) and their self-Coulomb energy, and (iii) an unknown functional already mentioned, called the exchange–$^o$correlation functional. We will begin by following the usual convention of writing the exchange-correlation functional $E_{xc}$ as the sum of an exchange contribution and a correlation contribution:

$$E_{xc} = E_x + E_c \qquad (1)$$

Each term in equation (1) can be a functional of the total electron density $\rho$ or—in a spin-polarized case—of the two spin densities $\rho_\sigma$, namely $\rho_\alpha$ for up-spin electrons and $\rho_\beta$ for down-spin electrons, where the total density is

$$\rho = \rho_\alpha + \rho_\beta \qquad (2)$$

The density functional can also be a functional of several other quantities; in particular we consider functionals that depend on one or more of the following additional variables:

- the local dimensionless reduced spin-density gradients
- the local spin-labeled kinetic energy densities $\tau_\sigma$
- the HF energy density.

We emphasize that here—and usually when we mention HF exchange— we are referring to the HF method for calculating the exchange energy from the Kohn–Sham Slater determinant, not to the actual HF exchange energy, which is what one would get from the HF Slater determinant.

Note that the dependency on $\tau_\sigma$ or the HF energy density yields an orbital-dependent functional that depends on the occupied Kohn–Sham orbitals. Such a functional is still a density functional because the Kohn–Sham orbitals are functionals of the density [6]. In the present article we will not consider density functionals that depend on unoccupied orbitals (such functionals, sometimes called doubly hybrid functionals [7,8], are potentially more accurate than the functionals considered here, but they are also more expensive).

It is useful to provide definitions of the dimensionless reduced spin-density gradients and of the local spin-labeled kinetic energy densities. But we note that different authors define these quantities in different ways. In particular, two main conventions are commonly used: one is usually used by Perdew

$$s_\sigma = \frac{|\nabla \rho_\sigma|}{2\left(6\pi^2\right)^{1/3} \rho_\sigma^{4/3}} \qquad (3)$$

$$\tau_\sigma = \frac{1}{2} \sum_{i=1}^{n_\sigma} |\nabla \psi_{i\sigma}|^2 \qquad (4)$$

and the other is usually used by Becke

$$x_\sigma = \frac{|\nabla \rho_\sigma|}{\rho_\sigma^{4/3}} \qquad (5)$$

$$\tilde{\tau}_\sigma = \sum_{i=1}^{n_\sigma} |\nabla \psi_{i\sigma}|^2 \qquad (6)$$

These conventions differ only by numerical factors; therefore they only involve differences in how the formulas are written and how the axes are scaled in plots. However it is important to be aware of these differences, especially when functionals are analyzed or implemented into software. We recommend using different symbols for the two sets of definitions so it will be clear which is being used.

Before proceeding we comment further on the partition of the exchange-correlation functional into two terms in equation (1). This is very convenient and sometimes useful for understanding the structure of the unknown exact functional, but it is somewhat arbitrary since no physical result depends on $E_x$ or $E_c$ separately—only the sum has physical meaning [9,10]. Furthermore the language is different from the familiar language of WFT. In particular $E_c$ includes only what is called dynamical correlation in WFT, and $E_x$ includes both exchange and left-right correlation [6,11]. Some workers use scaling criteria [12] to distinguish

"exchange" from "correlation". However, approximate density functionals include some exchange effects in $E_c$ and some correlation in $E_x$ simply because the approximation is imperfect. Having recognized this we will use $E_x$ and $E_c$ simply to indicate the motivations for the functional forms without repeating the warning that our partition might not agree with that preferred by others.

It should be noted that essentially all exchange-correlation functionals have empirical elements. Some have fitted parameters, some have parameters inherited from previous work, some involve an experience-based selection of constraints or functional forms, and many have more than one of the various kinds of empiricism; and parameters are fit to both experimental data and fundamental ab initio data. There is no unique way to count empirical elements, so we do not include such counts. Furthermore, we think that the key issue is choice of functional form, not number of parameters. (For example, is a functional with a sixth order polynomial in the density gradient and an empirical amount of Hartree-Fock exchange less empirical than a functional with an eighth order polynomial in the density gradient and no Hartree-Fock exchange parameter?) In the literature, success of functionals based on fitting large number of parameters is sometimes attributed solely to that very fact, as if a large number of parameters is sufficient to obtain good results. This represents a fundamental misunderstanding. For a given functional form that does not contain the needed dependence on the critical values of the density, density gradient, or other variables, one cannot fit a broad set of data even if one increases the number of parameters to be arbitrarily large. Therefore a key issue is the choice of functional form, and that is what we discuss next.

### 2(a) Local approximations: LSDA, GGA, NGA, and meta-GGA

Because energy densities that depend only on the spin-densities, their gradients, and the spin-labeled kinetic energy densities depend only on the local values of these variables at a given point in space they are called local functionals. The oldest density functionals, dating back to the original Kohn-Sham article [2] with roots in the pre-DFT literature [13-16], are local and are based on the uniform electron gas (UEG), which is a fictional system with a constant electron density generated by a smeared out positive background charge (not nuclei). Such approximations are sometimes called UEG approximations or free-electron-gas approximations, but more often they are called the local density approximation (LDA, when one is discussing only closed-shell singlets) or called the local spin-density approximation (LSDA, in the general case) to denote that they only depend on $\rho_\sigma$, not on $s_\sigma$, $\tau_\sigma$, or the HF exchange density. (The reader should not confuse the "local density" functionals that depend only on the local densities, called LDA or LSDA, with the more general "local" functionals (sometimes called "semilocal" functionals, especially in the physics literature) that also depend on other local variables. The reader should also note that in the early literature functionals now called "local" or "semilocal" were sometimes called "nonlocal," a confusing practice that now seems to have disappeared.)

The exchange energy of the UEG for a spin-polarized system has a simple mathematical expression (see the appendix for a description of the differences between the spin-polarized and spin-unpolarized case) [2,15]:

$$E_{\mathrm{x}} \equiv \sum_\sigma \int d\mathbf{r} \left\{ -\frac{3}{2} \left( \frac{3}{4\pi} \right)^{1/3} \rho_\sigma^{4/3} \right\}. \qquad (7)$$

Equation (7) may be called the Gáspár–Kohn–Sham (GKS) approximation for exchange (we warn the reader that "GKS" should not be confused with other uses of the same acronym, such as to stand for generalized Kohn-Sham).

The correlation energy of the UEG is usually based on parameterizations of quantum Monte Carlo calculations. UEG correlation functionals include the work of Vosko, Wilk, and Nusair (VWN) [17] with both their formula 3 and formula 5 being used in current software. More recent work is that Perdew and Wang in 1991 and 1992 [18], which is denoted PW92 when one is discussing UEG correlation. The differences in the mathematical forms of these parameterizations are sometimes significant, but the numerical results that they provide are very similar in most cases. A review of the UEG correlation energy is available [19] and further theoretical progress has also been achieved [20].

Early attempts to go beyond the UEG involved gradient expansions (i.e., a power series in a suitable function of the density gradient); however, it was noted early on [21] that "Opinion about the usefulness of including the lowest gradient correction in $E_{\mathrm{xc}}[\rho]$ in real condensed matter systems (in which the density gradients are *not* small) is divided." A brief summary of the unsatisfactory properties of the truncated gradient approximation is available [22]. In light of this situation, a more general approach was therefore widely adopted in which density functional approximations to the exchange energy $(E_\text{x})$ beyond LSDA are written as functionals of the spin-density distributions, $\rho_\sigma$, and their reduced gradients, $s_\sigma$, as:

$$E_\text{x} = \sum_\sigma \int d\mathbf{r} \, \Gamma_{\text{x}\sigma}\left(\rho_\sigma, s_\sigma\right). \tag{8}$$

but the leading terms in the expansion are not required to be correct. In generalized gradient approximations (GGAs) for exchange [23], $\Gamma_{\text{x}\sigma}$ is factored by a scaling argument into:

$$\Gamma_{\text{x}\sigma}\left(\rho_\sigma, \nabla\rho_\sigma\right) = G_{\text{x}\sigma}\left(\rho_\sigma\right) F_{\text{x}\sigma}\left(s_\sigma^2\right), \tag{9}$$

where $G_{\text{x}\sigma}$ is usually taken as the LSDA limit of equation (7). If we set $F_{\text{x}\sigma} = 1$, we get back the local spin-density approximation; and therefore $F_{\text{x}\sigma}$ (which is usually greater than unity) is usually called the GGA enhancement factor, which by definition depends only on the dimensionless reduced gradient. Various approximations have been proposed for the enhancement factor; Langreth and Mehl [24], Perdew and Wang [23], and Becke [25] carried out early work in this area. Resulting popular GGAs include the asymptotically correct exchange functional of Becke (B88) [26], the constraint-selection-based PBE functional of Perdew and coworkers [27], whose exchange portion is very similar to the earlier empirical Xαβγ functional of Becke, here called B86, and the optimized exchange functional of Handy and Cohen (OptX) [28]. GGAs also involve correlation functionals depending on $\rho_\sigma$ and $s_\sigma$ (for example, combining B86, B88, or OptX with the LYP GGA [29] for correlation yields respectively the B86LYP, BLYP, and OLYP functionals), and they often, but not always, tend to the UEG limit when $s_\sigma$ goes to zero. Our recent second-order GGA functionals SOGGA [30] and SOGGA11 [31] tend to the UEG limit in the correct way not just at $s_\sigma = 0$, but also to the analytically known leading term in an expansion in $s_\sigma$ for a nearly uniform electron gas. This term is second order in $s_\sigma$.

A major drawback of the GGA functional form is that it cannot satisfy all the theoretical constraints of the exact functional, and it has been found in practice that a single GGA cannot provide good accuracy for all databases of interest. One of the most significant consequences of this fact is that no GGA in the form of equation (9) has been able to provide good atomization energies for molecules and also good performance for lattice parameters of solids. An alternative (slightly weaker) statement of this fact is that "No single GGA can describe with high accuracy the properties of both solids (surface energies and lattice constants) and molecules (total and atomization energies)." [32] Theoretical considerations and results with simple functional forms suggested [25,27,30,33-35]mad that a GGA that provides good atomization energies needs a second order coefficient in the exchange functional (i.e., the coefficient of $s_\sigma^2$ in a Taylor series for $F_{x\sigma}$) that is about twice as large as the exact coefficient calculated from the gradient expansion in the slowly varying limit. However using the correct value for this second order coefficient was found to be one of the key ingredients necessary to obtain correct lattice parameters in solids, and this led to the creation of functionals such as PBEsol [33] that intentionally sacrifice performance for molecular energies to empirically improve the lattice constants by enforcing the second-order constraint on exchange. One must, however, be cautious about this kind of reasoning because it has often been based on studies with very restricted functional forms for GGAs. If a GGA has a limited functional form with only a few parameters, changing the functional form to change any one property or to fit any one constraint makes a global change in the functional, and one cannot be sure about which aspect of the resulting change in the functional is most responsible for the observed change in predictions. For example, the very flexible SOGGA11 functional [31] recovers the exact second order coefficients in both exchange [36] and correlation [37] and is parametrized for broad accuracy for chemistry [31]; however its performance for lattice constants is poor [38], clearly showing the fact that there is more involved than simply changing the second-order coefficient.

Although various arguments can be used (though nonuniquely) to define exact exchange in the context of DFT [10,39], one can also argue that "exchange and correlation need not ... be separated in DFT" [40]. Progress beyond the conventional GGA, but with the same ingredients, was achieved recently by the introduction [41] of a new functional that makes use of the same ingredients as GGA functionals (which are the spin-densities and their gradients), but uses a more general functional form in which an exchange-type term is not separable as is equation (9). This new kind of functional approximation includes both exchange and correlation in a nonseparable way by a new functional type that has the form of a non-separable gradient enhancement of UEG exchange; it also includes a more conventional correlation term. Functionals built this way are called nonseparable gradient approximations (NGAs), and the first specific realization is called N12 [41].

To go beyond the generalized gradient approximation and the non-separable gradient approximation, one needs to add more ingredients to the functional form, and the most popular way to do so has been to add either the second derivatives (Laplacian) of the spin densities or the spin-labeled noninteracting kinetic energy densities. Functionals incorporating either of these ingredients receive the label meta. Numerical instabilities linked to the use of the Laplacian in functional approximations, however, are one reason that made the noninteracting kinetic energy density the preferred ingredient for such approximations, which are usually built on GGAs and called meta-GGA functionals. We have also built a meta functional on an NGA; the resulting meta-NGA is discussed in section 2(c).

## 2(b) *Hybrid functionals and range separation*

As mentioned above, the Kohn-Sham equations include the Coulomb interaction of the electron density with the nuclei, and they also include the classical Coulomb self-energy of the electronic charge distribution. The classical self-interaction energy is nonlocal because the energy density at a given point involves an integration over all space. The exact exchange–correlation energy must also be nonlocal to correct the classical approximation in the self-energy [42] Hybrid GGAs replace a percentage of the local exchange by HF exchange. The motivation for this is the nature of the error in the classical self-energy, namely that the classical approximation includes the interaction of the entire electron distribution with itself by Coulomb's law without recognizing that the parts of an electron distribution corresponding to the same electron do not interact with one another. In DFT, this spurious self-interaction muSt be removed by the exchange-correlation functional. A local functional cannot completely remove this self-interaction so the unknown exact functional must be nonlocal. In WFT this spurious self-interaction is removed by antisymmetrization, and for a single Slater determinant, this results in the Hartree-Fock exchange energy. An energy density that depends on HF exchange can therefore provide a better approximate functional by removing some of the self-interaction.

Hybrid functionals started to became very popular after the broad success of the B3PW91 [26,43] functional and its even more successful close cousin B3LYP [26,29,44]. Following the most recent developments in density functionals, hybrid functionals that have a constant percentage of HF exchange ("constant" meaning independent of density, density gradient, interelectronic distance, or position in space) are now called global-hybrid functionals. Although more advanced and accurate functionals have been developed throughout the years, global-hybrid GGAs still remain among the most used functionals for chemical applications, probably because of their broad availability in popular quantum chemistry computer programs and their well earned user familiarity. As we recently noted [45]: "the B3LYP global-hybrid GGA functional is still the most popular functional in most areas of quantum chemistry, despite its known shortcomings. In part, this is true because DFT is now widely used by nonspecialists, and the early successes of B3LYP gave it a good reputation and made it hard to displace even with better performing global-hybrid GGAs". Not only are there better performing global-hybrid GGAs, there are also hybrid meta-GGAs of various types that perform even better. Nevertheless, for reasons of simplicity, one may sometimes prefer a global hybrid GGA, and we have recently optimized such a functional against a broad range of chemical properties; the resulting functional is called SOGGA11-X [45].

A more recent development in the creation of new hybrid functionals is represented by range-separation [46,47]; the basic idea of this approach is that the interelectronic Coulomb operator can be split into a short–range (SR) part and a long–range (LR) part. This effort is usually achieved by using an operator such as [46]:

$$\frac{1}{r_{12}} = \underbrace{\frac{\mathrm{erfc}\left(\omega r_{12}\right)}{r_{12}}}_{\mathrm{SR}} + \underbrace{\frac{\mathrm{erf}\left(\omega r_{12}\right)}{r_{12}}}_{\mathrm{LR}} \qquad (10)$$

where $r_{12}$ is the interelectronic distance, and the error function is used because it allows a simple calculation of the integrals. However, in principle, any other separation can be used, and some other functions—such as the Yukawa potential—although less common—have been employed, leading to similar results [48-50].

The most popular kind of range-separated hybrids are called long-range-corrected (LC) functionals [46,47,51,52]; they use 100% HF exchange in the LR limit and a smaller value, usually between 0 and 50%, in the SR limit. In the LR limit, 100% HF exchange is used to compensate part of the self–interaction error of DFT, since HF leads to an effective potential that has the correct asymptotic behavior. A closely related range-separation strategy is that employed by the CAM-B3LYP functional [53], which has 19% nonlocal exchange in the SR limit and 65% in the LR limit.

The opposite approach to LC is used in the so-called screened exchange hybrid functionals, such as the HSE06 functional (sometimes called HSE) [54,55] and our recent N12-SX [56] and MN12-SX [56] functionals. This approach uses a finite amount of HF exchange at SR, and but none in the LR limit, in order to cut the computational cost of nonlocal exchange integrals for extended systems, while at the same time—in principle— retaining the good performance features of global-hybrid functionals for most chemical properties. A second possible reason to use screened exchange is the argument that dielectric and correlation effects screen long-range exchange. We note that the savings in computer time as compared to hybrid functionals that do not screen the exchange depend strongly on the software, but in some cases "it reduces significantly" [57], for example due to reducing the number of $k$ points needed for calculations with periodic boundary conditions [58]. Nevertheless, "Despite extensive efforts towards computationally efficient implementations, HSE is still rather more expensive than" local functionals [59].

Range separation can also be used to mix two local functionals so the percentage of HF exchange is zero for both SR and LR [60]. Note that a range-separated local functional is not separable in the sense of eq. (9). Thus combining two meta-GGAs via range separation yields a meta-NGA.

A word is in order here about HF exchange and multireference systems. Multireference systems are most easily defined in the language of WFT. Multireference systems are those that cannot, even to zero order, be well described by a wave function in the form of a single Slater determinant, which is the HF approximation. Prominent examples include diradicals and systems with stretched bonds. The variational energy lowering (as compared to that obtained with a single Slater determinant) when one uses the smallest number of determinants that provide a good zero-order description is sometimes called static correlation energy or—in some cases—left-right correlation energy (although one should be careful to note that mixing two or more determinants corresponding to different electron configurations in WFT inevitably brings in some dynamical correlation as well). Thus static correlation energy is a special type of error in the HF approximation. Hybrid functionals, by using HF exchange, build in this error, whereas local functionals need not have this kind of error, and in fact it is an advantage of DFT that local exchange brings in some left-right correlation energy without the expense of a multiconfigurational wave function [11]. But, as mentioned above, a local functional cannot completely remove the spurious self-interaction of the classical electron repulsion in the Kohn-Sham equations. Thus in practice there is always a tradeoff, with zero or low HF exchange leading to a smaller static correlation error, but higher self-interaction error, and high HF exchange having higher static correlation error and lower self-interaction error.

### 2(*c*) The Minnesota meta functionals

Several meta functionals developed in Minnesota in 2005 and later have been given names of the form M*yz* or M*yz*-suffix, where *yz* denotes the year 20*yz*, and M denotes Minnesota or meta. Such functionals have been called Minnesota functionals, and they are our functionals with the broadest accuracy. When we develop functionals that do not include a dependence on kinetic energy density, which we consider to be an important ingredient in a broadly applicable density functional, it is usually to demonstrate what can be done with a restricted set of ingredients, which is a demonstration that is of interest both for fundamental understanding and for ease of implementation, but nevertheless we cannot expect the highest possible accuracy if we forego dependence on kinetic energy density. The Minnesota functional family consists of one meta-GGA (M06-L), two meta-NGAs (M11-L and MN12-L), seven global-hybrid meta-GGAs (M05, M05-2X, M06-HF, M06, M06-2X, M08-HX, and M08-SO), one range-separated hybrid meta-GGA (M11), and one screened exchange hybrid meta-NGA (MN12-SX).

These functionals are all parameterized against a broad range of chemical data. The success (or lack of success) of a given one of these functionals depends on the design of (or choice of) an appropriate functional form as well as the parametrization strategy (what data to use, how to weight the various items of data, and how to optimize linear and nonlinear parameters). In some cases the desired behavior can be designed in, whereas in other cases the "results show how a flexible functional form can lead to the 'discovery' of a desired behavior of a functional." [61] Next we explain the Minnesota functionals in chronological order.

- M05 family [62,63]: The first meta-GGA functional to be named Minnesota functional dates back to 2005, when we first used a flexible functional form to optimize a meta-GGA functional, called M05 [62], on a large number of databases representing important chemical properties. We found that the M05 functional was able to give better balanced accuracy for chemical reaction barrier heights and bond energies in molecules containing metals than any previously available functional. We also optimized a related functional, called M05-2X [63], with a percentage $X$ of HF exchange that is twice as high. As far as energetic properties are concerned, M05-2X was found to perform much worse than M05 for many systems containing transition metals, but is better for almost all other systems, the exception being systems with high multireference character; however the increase of $X$ often makes geometries and vibrations slightly worse. The conclusions about applicability to multireference systems can be understood in part from the discussion above of the static correlation error brought in by Hartree–Fock exchange, and they will be re-examined in the present survey using new databases that are more extended and reorganized as compared to those used in the original M05 [62] and M05-2X [63] publications.

- M06 family [64-66]: The M06 family of functionals is composed of four functionals that have similar functional forms for the DFT part, but each has parameters optimized to be used with a different percentage of HF exchange. The four functionals are: M06-L [64], a local functional (no HF exchange); M06 [65], a global-hybrid meta-GGA with 27% of HF exchange, leading to a well-balanced functional for overall good performance for chemistry; M06-2X [65], a global-hybrid meta-GGA with 54% HF exchange, for top-level across-the-board performances in all areas of chemistry including thermochemistry and reaction kinetics, but excluding multireference system such as many systems containing transition metals; and M06-HF [66], a global-hybrid meta-GGA with 100% HF exchange, suitable for calculation of spectroscopic properties of charge-transfer transitions, where elimination of self-interaction error is of paramount importance. Although it was formerly believed that one could not design a generally useful functional with 100% HF exchange, M06-HF disproved this by achieving overall performances for chemistry better than the popular B3LYP functional.

- M08 family [67]: The success of the M06 family of functionals is largely documented in previous studies, and is confirmed by their popularity. Noting that we were very close to the limit that the M05/M06 functional form could provide, in 2008 we made an attempt to improve it [67]. This led to a new global-hybrid meta-GGA functional called M08-HX where HX stands for *Hi*gh-HF e*X*change; it uses 52.23% of nonlocal exchange. M08-HX is an improved version of the previous high-HF exchange Minnesota functionals M05-2X and M06-2X. The M08-HX functional performs similarly to the M06-2X functional, but it uses a much cleaner functional form for both the exchange and the correlation. This functional form later became the base for the future Minnesota functionals, such as the M11 family. Following the development of the SOGGA functional, we also introduced a hybrid meta-GGA functional, called M08-SO, that is correct through second order in both the exchange and the correlation by enforcing the second order constraint on the M08 functional form. M08-SO is a high-exchange functional (with 56.79% of HF exchange), and its performance is very similar for many (but not all) properties to that of M08-HX. The M08-SO functional was the first Minnesota family functional to be correct through second order. (SOGGA is not called a Minnesota functional because it does not depend on the kinetic energy density.) Despite having improved the functional form—especially in the terms that depend on the kinetic energy density—and having eliminated a problematic term from the VS98 exchange to reduce the grid sensitivity, both M08 functionals still require ultra-fine quality integration grids, mainly because of the sensitivity coming from the correlation part. The grid sensitivity and the related occasional convergence problems within self-consistent-field calculations are present—to different extents—in all meta-GGA functionals, including the M11 family discussed below and including meta-GGAs from other groups. This represents an open problem in DFT development, still to be solved. The grid sensitivity should not, however, be overemphasized. It has not been a hindrance to using the functionals for practical work, although the need for a finer grid can slightly raise the cost of a calculation.

- M11 family [60,68]: In the recent 2011 generation of Minnesota functionals we introduced the range-separation feature into our already very successful functional form. The M11 functional [68] is entirely based on the M08 meta-GGA functional form for the DFT part, but it uses the long-range correction scheme in the way employed by Chai and Head-Gordon [69] for a GGA, with the percentage of HF exchange in M11 varying from 42.8% at SR to 100% at LR. This means that the percentage of HF exchange in M11 is always greater than or equal to 42.8%, and this puts it in the class of the high-$X$ functionals (such as M05-2X, M06-2X, M08-HX and M08-SO), but the fact that it has 100% of HF exchange in the LR provides it with the chief advantage of M06-HF. Range-separation in the M11-L functional [60] was introduced at the local level only, by introducing a new strategy called dual-range exchange. The dual-range exchange strategy of M11-L uses two different local functionals, one for SR and one for LR, and each local functional is a flexible meta-GGA. The meta-GGA forms are parametrized using databases of accurate reference data, as has been done for all Minnesota functionals. The main differences between the dual-range approach (as in M11-L), the long-range-corrected approach (as in M11), and the screened exchange approach (as in HSE06) are schematically illustrated in Figure 1. The performance of the M11-L functional is much better than its predecessor, M06-L, for many classes of interaction, and this often makes M11-L a potential substitute even for the M05 and M06 hybrid functionals, avoiding the computational cost of the expensive HF exchange. In summary, in the M11 family, M11-L in principle replaces M06-L, M11 in principle replaces M06-2X, M08-HX, M08-SO, and M06-HF, and M06 is in principle replaced by either M11 or M11-L, thereby bringing us closer to a single universal functional, which is still an unmet challenge for long-term research. This is illustrated in the genealogy tree of Figure 2. We note though that the "in principle" qualification should not be taken lightly; at this stage there is more experience using the older functionals, and one should be cautious when switching to the less extensively vetted newer ones. This caution is particularly true when it comes to transition metals and other metals. Although we have placed more emphasis in recent work on systems containing metals and systems with multireference character, our training sets are still not good enough in this regard. This is especially true because there is more than one kind of multireference character, and we are still sorting out the issues involved in designing density functionals that can treat metal-containing compounds and systems with multireference character reliably.

- MN12 family [56,70]: The MN12 family includes MN12-L [70] and MN12-SX [56]. These functionals both build on the N12 nonseparable functional form discussed above, with MN12-L adding kinetic energy density and MN12-SX adding both kinetic energy density and screened exchange, which is also explained above. The MN12-L functional is particularly accurate for atomization energies, ionization potentials, barrier heights, noncovalent interactions, isomerization energies of large molecules, and solid-state lattice constants and cohesive energies [70]. MN12-SX is better than MN12-L for main-group atomization energies, electron affinities, proton affinities, alkyl bond dissociation energies, hydrocarbon energetics, $\pi$-system thermochemistry, barrier heights of all kinds, noncovalent interaction energies, difficult cases, atomic energies, main-group nonhydrogenic bond lengths, and semiconductor band gaps, and it has the best overall performance on the CE345 database of any functional we have considered, but we still require experience on a broad range of applications to say how useful it will be in the long run.

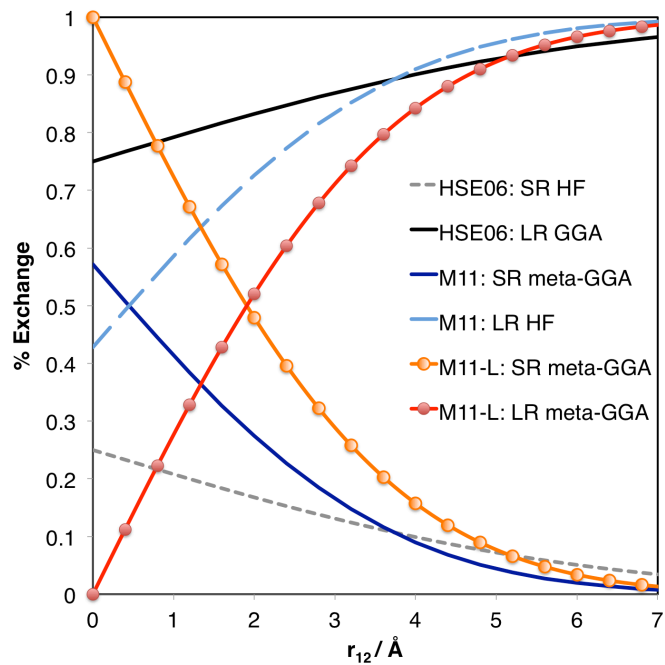

Figure 1. Difference between three kinds of range-separation: screened exchange (HSE06, shades of black), long-range corrected (M11, shades of blue), and dual-range (M11-L, shades of red) exchange functionals.

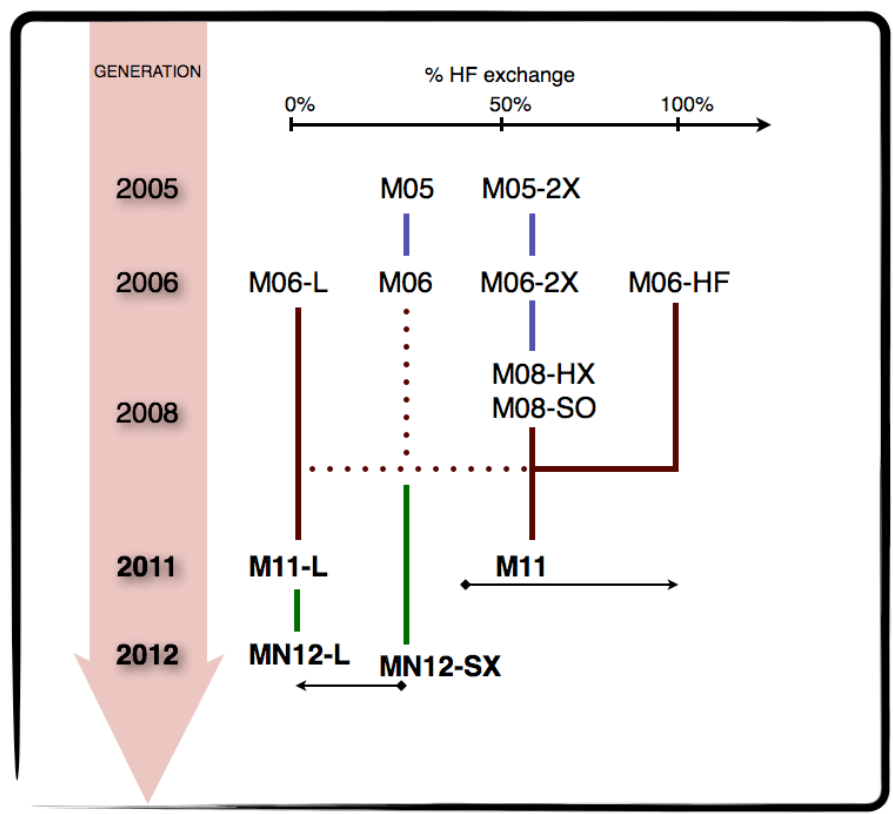

Figure 2. Genealogy tree of the Minnesota families of meta density functionals (note that the SOGGA, SOGGA11, SOGGA11-X, and N12 density functionals are not included in this diagram because they do not include kinetic energy density—i.e., they are not meta functionals—, and functionals developed before 2005 are not included either).

In classifying functionals, many workers use the Jacob's ladder scheme of classification, first introduced by Perdew and Schmidt [22]. The functional approximations described in subsection (*a*) of this section are all on rungs 1–3 of Jacob's ladder, hybrid functionals are on rung 4, M06-L, M11-L, and MN12-L are on rung 3, and the other Minnesota functionals are on 4. The ladder scheme does not distinguish hybrid GGAs from hybrid meta-GGAs, nor conventional GGAs from an NGA, nor does it distinguish range-separated hybrids from global-hybrids or dual-range local functionals from single-range ones; however these differences are very important.

## 2(*d*) Functionals with nonlocal correlation

For exchange, the results presented in the present article include both local (LSDA, GGA, NGA, meta-GGA, and meta-NGA) approximations and nonlocal ones (various hybrid mixtures of HF exchange). One may also consider nonlocal correlation, and this may be done with our without considering unoccupied orbitals. In the present article, though, we do not consider such functionals. Certainly they are more complicated, and usually they are much more expensive to compute. Furthermore, they are still being heavily vetted, and no firm conclusion has emerged. Some examples of functionals with nonlocal correlation are MC3BB [7], MCG3/MPWB [71], vdW-DF [72], TPSS/CCSD(T) [73], vdW-DF2 [74], sc-NEVPT2-srDFT [75], optB88-vdW [76], optB86b-vdW [77], VV10 [78], PWPB95-D3 [79], and LC-VV10 [80]. These functionals deserve a separate review.

## 2(*e*) Functionals including molecular mechanics

A large industry has developed for adding post-SCF molecular mechanics terms to various density functionals. The motivation is to add long-range dispersion forces, which are missing in functionals with local (LSDA, GGA, or meta-GGA) correlation. This work deserves some comments. For a start, we note that polar molecules have significant electrostatic and inductive forces, and the dispersion forces need not dominate the noncovalent attraction. Many density functionals predict reasonably accurate multipole moments and polarizabilities, so they do quite well for electrostatic and induction forces. Attention has therefore centered on dispersion forces. We note though that "dispersion" is an ambiguous term. Its original usage was to refer to long-range forces treated by perturbation theory of separated subsystems with negligible overlap; the formalism involved a sum over excited states, and the name invokes the Kronig-Kramers dispersion relations that characterize the spectrum of any molecule because some of the same quantities that appear in the perturbation expressions for long-range forces also appear in those dispersion relations. What is not always clear when one discusses dispersion interactions is that the assumption of negligible overlap is by no means true at the minimum energy geometries of van der Waals molecules. At such configurations, since they are energy minima, the gradient of the repulsive forces has the same magnitude as the gradient of the attractive force (they are equal and opposite). But the repulsive forces between nonpolar molecules and atoms are due to exchange repulsion, which results from overlap of the orbitals on the two subsystems. When overlap is present, as at van der Waals minima, we prefer to speak of dispersion-like forces or medium-range correlation energy, saving the terms "long range" and "dispersion" for the essentially overlapless region of interaction. Although functionals with local correlation (LSDA, GGA, and meta-GGA) cannot predict the long-range induced dipole–induced dipole dispersion forces, they *can* predict medium-range attractive noncovalent interaction at van der Waals distances where overlap is not negligible. However, many available functionals underestimate these forces or even predict net repulsion at van der Waals distances. That is the motivation for adding molecular mechanics attraction that decays at long range as $R^{-6}$, where $R$ is the distance between subsystems. An example of how molecular mechanics "dispersion" cancels the overly repulsive nature of intermolecular interactions provided by the popular B3LYP functional has been provided by Acosta-Silva et al. [81]

One must modify the $R^{-6}$ functional form at medium range for two reasons: (i) attractive terms of order $R^{-8}$ and $R^{-10}$, even $R^{-7}$, become significant at medium range, so $R^{-6}$ underestimates the undamped force, and (ii) orbital overlap makes the multipole expansion leading to these powers invalid, and so the expansions must be damped to avoid overestimating the noncovalent attraction.

Molecular mechanics dispersion involves first a choice of functional form, then parameters or approximations to predict the coefficient of $R^{-n}$ terms ($n \geq 6$) and their dependence on molecular geometry and bonding pattern, and finally parameters in the damping function(s). Because there is no unique definition of the dispersion-like contribution once overlap becomes significant, and because the parametrization is making up for errors in the density functional without added terms, the parametrization cannot be validated independently of the non-dispersion-like interactions, and the parameters depend strongly on which underlying functional is involved. In most cases, the parametrization is made to be independent of charge state, oxidation state, partial atomic charges, hybridization, and bonding pattern, which can be a serious approximation, and if $R^{-n}$ terms are underdamped, they can cause large errors in some cases. Grimme has studied these issues most carefully; he eventually became unsatisfied with his first two rounds of parametrization and all other molecular mechanics approaches to the dispersion problem, so he devised a set of "D3" third-generation functional forms and parametrizations, which are the most complete attempt to minimize these difficulties [82] Even there though, it was found that a special reparameterization was required to treat ionic surfaces [83], and then the D3 method was further improved by the D3(BJ) method [84]. Similarly to D3 and D3(BJ), the TS-vdW molecular mechanics formalism also involves a parametrized damping function which is be reparameterized for each functional [85] because the "dispersion" terms include not only dispersion-like contributions to the interaction energy but also molecular mechanics corrections for systematic errors in the underlying xc functional.

Molecular mechanics treatments of dispersion-like interactions also have the disadvantage that when one adds pairwise atom-centered dispersion terms, there is no accounting for whether these interactions are screened by other parts of the system that may lie between the two interacting atoms.

Despite these concerns, we do include two functionals with post-SCF molecular mechanics "dispersion" terms in the tests reported here, but that is only a small subset of the many available parametrizations. Our own attitude is to prefer density functionals that predict the medium-range attractive noncovalent interactions as part of the density-dependent self-consistently used functional itself, and the Minnesota functionals as a group tend to do a better job of this than other available functionals with local correlation. Karton et al. [86] devised a damped dispersion correction and empirically determined a scale factor by which it should be multiplied when used with various underlying functions. For example, for B97-3, the scale factor is 0.90, indicating that only 10% of the dispersion-like interaction terms were needed (on average), and for BLYP the factor was 1.20, indicating not only the absence of dispersion-like interactions, but a systematic overestimation of medium-range repulsive forces that need to be compensated by a factor great than unity. The scale factors for some other functionals considered in this article were 1.10 for B3PW91 and HCTH407, 1.05 for B3LYP, 1.00 for TPSS, 0.90 for TPSSh and TPSS1KCIS, 0.75 for PBE, 0.70 for PBE0, and 0.765 for BMK. However, the required factor was only 0.50 for PW6B95 (a precursor of M05), 0.25 for M06, 0.20 for M06-L, and 0.06 for M06-2X. In other words, M06-2X already includes, on average 94% of the empirically needed dispersion-like interactions. Similarly, when Grimme and coworkers added dispersion-like terms to the M05-2X, M06-L, M06-HF, M06, and M06-2X functionals, their parameter values indicate a need for correction mainly at large $R$, not at medium $R$ [87]. A third example like this is in the parametrization of the TS-vdW model where the "dispersion" terms where damped at a larger distance for M06 and M06-L than for six other studied functionals [85] because the Minnesota functionals already included more attractive noncovalent interactions at van der Waals distances. The results presented below show that two local functionals developed in Minnesota (M11-L and MN12-L) are more accurate than M06-L for noncovalent complexation energies, and five nonlocal functionals developed in Minnesota (M06-2X, M08-HX, M08-SO, M11, and MN12-SX) are more accurate than M06 for noncovalent complexation energies. Undoubtedly there are situations involving very large molecules or condensed phases where it would be helpful to add a long-range correction onto any functional based on local correlation, but the effect is expected to be small in most cases for the Minnesota functionals, and we have not pursued that. An extensive review of WFT and DFT calculations on noncovalent interactions has been provided by Riley et al. [88]

## 3. Computational aspects

There are a large number of programs that can perform DFT calculations, and the Minnesota functionals are included in several of them. All the results presented in this article were calculated using a locally modified version [89] of *Gaussian 09* [90]. The 2011 generation of Minnesota functionals is also implemented, at present, in the following programs: *Q-Chem* (as of version 4.0) [91], GAMESS (as of release R1 2012) [92], and *NWChem* (as a patch to version 6.1, and to be fully included in the next major release) [93]. The older Minnesota functionals are present in even more software, while the implementation of the 2012 generation is currently in progress; more details of the implementation of Minnesota functionals are given at http://comp.chem.umn.edu/info/DFT.htm.

Meta-GGA functionals are in general more sensitive to the integration grid than GGA functionals, and therefore they usually require a finer integration grid than the default of most popular programs. In the present paper though, all calculations, whether with or without kinetic energy density, were performed using the ultrafine ("99,590") Lebedev grid of the *Gaussian09* program [90]. In addition we always allow symmetry breaking in the orbitals of the Slater determinant in order to converge to the stable broken-symmetry solution (through the STABLE=OPT *Gaussian* keyword).

Another computational issue is basis set superposition error (BSSE). We do not include a counterpoise correction for such errors in the present work. The reasons for this are that we have found that such corrections do not necessarily improve the accuracy and there is no generally accepted scheme for applying such corrections to all problems of interest, for example, ternary interactions, transition state barrier heights, or condensed-phase problems. Our goal is to evaluate methods that yield useful results without such corrections.

## 4. Databases

Many databases have been collected and used throughout the years by our group for applications and development of DFT, and this effort is still an ongoing process with (especially) further work being carried out to develop more comprehensive databases for metal-ligand bond energies. The paper presenting the M06 and M06-2X functionals [65] summarized our most important databases as of 2006, and the present article summaries an important subset of our data as of 2012. In particular we summarize the current status of many of our most widely used databases, including in a few cases new data added for the first time here.

Each of our databases represents one or another particular class of properties—such as atomization energies, reaction barriers, lattice constants, band gaps, etc. Each database is dubbed with an acronym representing the particular property considered (e.g., MGAE for main group atomization energies), followed by the number of data in the subset (e.g., 109 data), and eventually—if there has been more than one version—by the last two digits of the year of latest revision of the database (e.g. 11 for 2011); if the year is not specified, it means that the database is at its first generation, and no revision of its data has ever been made (therefore its year is that of the corresponding reference).

We have combined several of our databases into four comprehensive databases, representing energetic and structural properties for chemistry and physics. The first database is called CE345 (chemistry energetic database with 345 data); the second database is called PE39 (physics energetic database with 39 data); the third database is called CS20 (chemistry structural database with 20 data); finally, the fourth database is called PS47 (physics structural database with 47 data). These comprehensive databases are the main subject of the present review, and we have made them available through a webpage called

http://comp.chem.umn.edu/db. In the next sections we will use these databases to assess the performance of a large number of density functionals, including the Minnesota functionals.

All energetic data in the databases are Born-Oppenheimer potential energy differences without zero point energy or thermal vibrational, rotational, or translational contributions.

Although the comprehensive databases were put together by combining data from other databases, it is easiest to understand their structure by dissecting them rather than discussing their assembly. The comprehensive databases may be considered level 1 of a hierarchy. Each of these level-1 databases contains nonoverlapping level-2 subdatabases, where "nonoverlapping" means that each datum appears in one and only one subdatabase. These level-2 databases will be called the primary databases in the present article. A list of the primary databases, together with the reference or references for the latest version can be found in Table 1, while a graphical summary is available in Figure 3.

Note that all comparisons to the CE345 reference data in Table 1 are carried out by means of single-point calculations at specified geometries, which are taken (as indicated in each row of the table) from experiment, HF, B97-D, or B3LYP calculations, Møller–Plesset second-order perturbation theory (MP2), quadratic configuration interaction with single and double excitations (QCISD), or multi-coefficient quadratic configuration interaction with single and double excitations, version 3 (MC-QCISD/3). In contrast, the comparisons to reference data in the PE39 database are carried out with consistently optimized geometries.

Table 1. Summary of the primary databases for chemistry and physics.

| Subset: | Description | Geometries | Reference(s) |
|---|---|---|---|
| | CE345 | | |
| MGAE109/11 | Main Group Atomization Energies | QCISD/MG3 | [45,63] |
| SRMBE13 | Single-Reference Metal Bond Energies | experiment | [60] |
| MRBE10 | Multireference Bond Energies | QCISD/MG3 and experiment | [60] |
| IsoL6/11 | Isomerization Energies of Large Molecules | B97-D/TZVP | [94] |
| IP21 | Ionization Potentials | QCISD/MG3 and experiment | [62,63,70,95-97] |
| EA13/03 | Electron Affinities | QCISD/MG3 | [62,63,95,96] |
| PA8/06 | Proton Affinities | MP2/6-31G(2df,p) | [98] |
| ABDE12 | Alkyl Bond Dissociation Energies | B3LYP/6-31G(d) | [31,63,64,99] |
| HC7/11 | Hydrocarbon Chemistry | MP2/6-311+G(d,p) | [31] |
| πTC13 | ThermoChemistry of π Systems | MP2/6-31+G(d,p) | [62,64,98] |
| HTBH38/08 | Hydrogen Transfers Barrier Heights | QCISD/MG3 | [63,71,100,101] |
| NHTBH38/08 | Non-Hydrogen Transfers Barrier Heights | QCISD/MG3 | [63,71,100,101] |
| NCCE31/05 | Noncovalent Complexation Energies | MC-QCISD/3 | [95,102] |
| DC9/12 | Difficult Cases | MP2/6-311+G(d,p) | [41] |
| AE17 | Atomic Energies | … | [65,103] |
| **PE39** | | | |
| SSCE8 | Solid-State Cohesive Energies | optimized | [30,104] |
| SBG31 | Semiconductor Band Gaps | optimized | [38] |
| **CS20** | | | |
| MGHBL9 | Main Group Hydrogenic Bond Lengths | optimized | [30] |
| MGNHBL11 | Main Group Non-Hydrogenic Bond Lengths | optimized | [30,41] |
| **PS47** | | | |
| MGLC4 | Main Group Lattice Constants | optimized | [30] |
| ILC5 | Ionic Lattice Constants | optimized | [30] |
| TMLC4 | Transition Metal Lattice Constants | optimized | [30] |
| SLC34 | Semiconductor Lattice Constants | optimized | [38] |

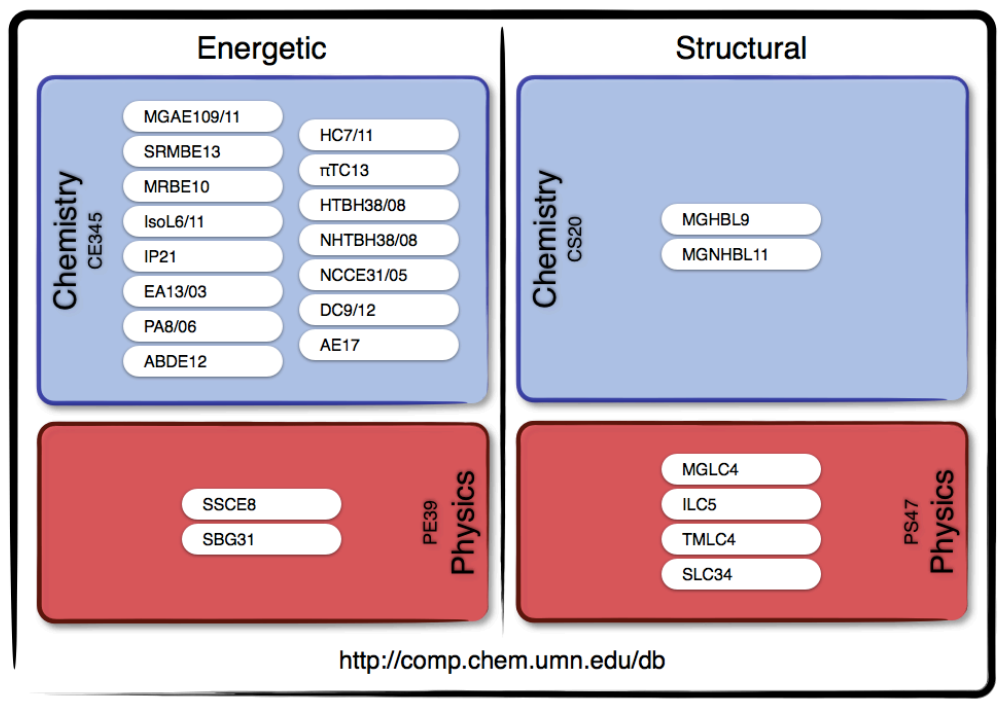

Figure 3. Schematic structure of the primary databases for chemistry and physics.

We will also consider two other kinds of databases, which we will call level-3 databases. Level-3 databases can be either subsets of level-2 databases (and in this cases we will call them secondary databases), or databases that have data from more than one primary database but are not simply the union of two or more other databases (and in this case we will call them analytical databases). Analytical databases are useful because we reorganized the data in order to allow a different kind of analysis of the performance of the density functionals.

The primary (level-2) databases are explained further in subsection *a*, while the secondary and analytical (level-3) databases are explained further in subsection *b*. Statistical conventions are explained in subsection *c*.

Although our databases are extensive and broad, they do not include all features present in previous assessments. Although DFT is increasingly being applied to excited-state molecular problems, the present article is limited to ground states and band gaps, although we do consider ionization potentials and electron affinities. The accuracy of DFT for molecular electronic excitation energies is assessed elsewhere [105-114]. A recent review [115] included 56 DFT assessment and validation articles for transition metals alone; as compared to the sum of the data in those assessments, transition metals are represented here with limited data, although some care was taken to make that data representative. (As mentioned above, our databases in this area are currently undergoing expansion.) A very large database for main-group chemistry, complementary to the present one, is provided by the GMTKN30 database of Grimme, which is collection of 30 subdatabases containing 841 relative energies and which has been thoroughly studied [87]. We have studied three of its subdatabases for reaction energies in subsequent work [116]. Mardirossian et al.[117] have also recently performed extensive systematic tests against multiple kinds of data. The extensive tests of density functionals by Rayne and Forest [118,119] are also noteworthy.

We note that databases of heats of formation (for example the G2/97 test set [120]) have been widely used for testing electronic structure theories, including DFT. Using such databases requires one to include vibrational zero point effects (for the heat of formation at 0 K) and thermal vibrational energies (for the heat of formation at 298 K). This raises issues of vibrational anharmonicity and allows for the possibility that errors in estimated vibrational contributions add to or cancel the errors in electronic energies. We prefer instead to use vibration-exclusive energies (also called zero-point-exclusive energies), which are differences in Born-Oppenheimer potential energy surfaces, and usually we carry out energetic testing at fixed geometries so all density functionals are evaluated at the same set of energies. As mentioned in the fourth paragraph of this section, this is the approach taken in the present article. This means that when the reference data are obtained from experiment, the vibrational and rotational effects are removed at the stage of developing the reference data.

### 4(a) The primary databases

The chemistry energetic database, CE345, has 15 nonoverlapping subsets of different properties; these are the primary chemistry energetic databases. Calculations are performed at fixed geometries, so all methods are compared at the same set of pre-determined geometries, which are specified in Table 1. The primary subsets that compose the chemistry energetic database, each representing a relevant property, are explained below:

- Main Group Atomization Energies (MGAE109/11): the main group atomization energy is composed of 109 atomization energies. It was introduced as an expansion of Database/3 [121] and Database/4 [122] and was first used for DFT development in the test set of the M05 functional [63]. The database was most recently updated in 2011 [45] by using more accurate reference data from W4 [123], W4.2 [123], W4.3 [123], and W4.4 [124] calculations. Geometries for all molecules are obtained at the QCISD/MG3 level of theory [125], and we used the MG3S basis set [96].

- Single-reference metal bond energies (SRMBE13): This database and the next one for multireference bond energies were recently formulated [31] and expanded [60] based on previous work on systems containing metals [126,127]. The first four data in this database are the bond dissociation energies for diatomic molecules containing metals that have positive MP2 binding energies, in particular: $Ag_2$, $CuAg$, $Cu_2$, and $Zr_2$. The next eight data are extracted from a previous database by choosing those that have B1 diagnostics [127] smaller than 10 kcal/mol, in order to include only data with single-reference character: AgH, CoH, $Cr(CH_3)^+$, $Cu(H_2O)^+$, FeH, LiCl, LiO and $V(CO)^+$. The final datum was added recently [60] by adding the bond energy for AlCl [128]. We used the def2-TZVP basis set [129,130] for the SRMBE13 database. Bond energies for all molecules in this database are equilibrium ones ($D_e$), obtained from the experimental bond energy in the ground vibrational state ($D_0$).

- Multireference bond energies (MRBE10): This database contains ten systems with high multireference character. Five data comes from a previous database of multireference metal bond energies (MRMBE5, [31]): $Cr_2$, $V_2$, $Ni(CH_2)^+$, $Fe(CO)_5$, and VS. Bond energies for these molecules are bond energy at equilibrium ($D_e$), obtained from the experimental bond energy in the ground vibrational state ($D_0$), as for the previous database. The remaining five data are taken from Karton et al. [128] and do not involve metals. All these data have high multireference character according to the %[(T)] diagnostic [123] (they also all have B1 diagnostics [127] greater than 10 kcal/mol), and the detailed dissociation reactions are: are: $B_2 \rightarrow 2B$, $O_3 \rightarrow O_2 + O$, $C_2 \rightarrow 2C$, $S_4 \rightarrow 2S_2$ and $Cl_2O \rightarrow Cl_2 + O$. Geometries for these five reactions are obtained at the QCISD/MG3 level of theory [125]. For the MRBE10 database we used the def2-TZVP basis set [129,130].

- Isomerization of large organic molecules (IsoL6/11): This database was introduced in order to include larger molecules in the training and performance evaluation of density functionals, and it is based on a larger database from Grimme, called IsoL22 [131]. However, since some of the reference data in the original IsoL22 are questionable, we recently [94] recalculated the reference energy for six of the smallest molecules in Grimme's database by using the accurate CCSD(T)-F12a/aug-cc-pVDZ method [132-136], and collect the results in the IsoL6/11 set. Geometries for this set are taken form the original work of Grimme [131] and are optimized at the B97-D/TZVP level [137,138]. For this database we used the MG3SXP basis set [67].

- Ionization potentials (IP21): The ionization potential (IP) database [70] contains data from six main group atoms (C, S, O, P, Si, and Cl), seven transition metal atoms (Cr, Cu, Mo, Pd, Rh, Ru, and Zn) and eight molecules (SH, $Cl_2$, OH, $O_2$, PH, $PH_2$, $S_2$, and FeC). Calculations on molecules involve separately optimized geometries for neutral and cations. The equilibrium bond length of FeC is obtained [97] from the experimental bond length [139] in the ground vibrational state, while geometries for all other molecules are obtained at the QCISD/MG3 [125] level of theory. We used the MG3S basis set [96] for the main group atoms and all molecules except FeC, for which we used the SDD+2fg [140] basis for Fe and the def2-QZVPP basis [130] for C; we used the cc-pVTZ-DK basis set [141] for the transition metal atoms. The scalar relativistic effects are included in the calculations of the seven transition metal atomic IPs by using the Douglas-Kroll-Hess (DKH) second-order scalar relativistic Hamiltonian [142-144], while they are included in the calculations of FeC with the SDD relativistic effective core potential [140].

- Electronic affinities (EA13/03): The electronic affinities (EA) database [62,63,95,96] contains six main group atoms (C, S, O, P, Si, and Cl) and seven molecules (SH, $Cl_2$, OH, $O_2$, PH, $PH_2$, and $S_2$). Reference data and geometries are obtained at the QCISD/MG3 level of theory [125]. Calculations on molecules involve separately optimized geometries from neutral and anions. For this database we used the MG3S basis set [96].

- Proton affinities (PA8/06): The proton affinities (PA) database [98] contains the proton affinities of the following small molecules: $NH_3$, $H_2O$, $C_2H_2$, $SiH_4$, $PH_3$, $H_2S$, HCl and $H_2$. As for the previous two sets calculations involve separately optimized geometries from neutral and charged (in this case protonated) molecules. Geometries are obtained at the MP2/6-31G(2df,p) level of theory. For this database we used the MG3S basis set [96].

- Alkyl bond dissociation energies (ABDE12): This database is a merger of ABDE4/05 and ABDEL8. ABDE4/05 contains four bond dissociation energies of small R–X organic molecules, with R = methyl and isopropyl, and X = $CH_3$ and $OCH_3$. $D_0$ values were taken from a paper by Izgorodina et al. [99], and we used the B3LYP/6-31G(d) zero-point vibrational energies scaled with a scale factor of 0.9806 to obtain our best estimate of the $D_e$ values in the database. For this database we used the MG3S basis set [96].

- Hydrocarbon chemistry (HC7/11): This database consists of seven cases of hydrocarbon data that are sensitive to medium-range correlation energy. HC7 is the combination of the HC5 database [145] with two isodesmic reactions (involving adamantane and bicycle [2.2.2]octane) that were singled out as difficult cases by Grimme. All geometries are obtained at the MP2/6-311+G(d,p) level of theory. The original reference data for this database was published in the original paper [65], and some inconsistencies in the reference data were corrected recently [31]. For this database we used the 6-311+G(2df,2p) basis set [146].

- Thermochemistry of $\pi$ systems ($\pi$TC13): This database containing $\pi$ systems [63,64] is composed of three isomeric energy differences between allene and propyne as well as higher homologs (which correspond to cumulenes and polyenes—this subset is called $\pi$IE3/06), five proton affinities of conjugated polyenes (PA-CP5/06) and five proton affinities of conjugated Schiff bases (PA-SB5/06). Geometries for all the molecules in this database are obtained at the MP2/6-31+G(d,p) level of theory, and we used the MG3S basis set [96].

- Hydrogen transfer barrier heights (HTBH38/08): This database contains 38 transition state barrier heights for 19 hydrogen transfer (HT) reactions, 18 of which involve radicals as reactant and product. Six reference data in the HTBH38 database were revised in 2008 [101]. All geometries are obtained at the QCISD/MG3 level of theory [125]. For this database we used the MG3S basis set [96]. All reactions in HTBH38/08 are isodesmic.

- Non-hydrogen transfer barrier heights (NHTBH38/08): The original version of this database was created in 2004 [71] by joining three older databases containing 38 transition state barrier heights for non-hydrogen-transfer (NHT) reactions. NHTBH38/08 contains 12 barrier heights for heavy-atom transfer reactions, 16 barrier heights for nucleophilic substitution (NS) reactions, and 10 barrier heights for non-NS unimolecular and association reactions. As for the previous case, geometries are obtained at the QCISD/MG3 level of theory [125]. 18 reference data in the NHTBH38 database were revised in 2008 [101]. For this database we used the MG3S basis set [96].

- Noncovalent complexation energies (NCCE31/05): Several databases have been developed in our group for various kinds of noncovalent interactions, and currently we use HB6/04 [102], CT7/04 [102], DI6/04 [102], WI7/05 [95], and PPS5/05 [95]. The geometries for the benzene dimers in the NCCE31/05 database are taken from Sinnokrot and Sherrill [147], while geometries for all other molecules in this database are optimized at the MC-QCISD/3 level [121,148]. For this database we used the MG3S basis set [96].

- Difficult cases (DC9/12): In this database of difficult cases for DFT we used the data from our previous DC10 database and omitted the datum for the atomization energy of ozone. The omission is to avoid a repetition in the database, since ozone is also present in the MRBE10 set. The /12 suffix was added to the name to avoid confusion with a database by Grimme which is called DC9, but contains different data. All geometries are obtained at the MP2/6-311+G(d,p) level of theory [125], while we used the MG3S basis set [96] for the calculations.

- Atomic energies (AE17): total atomic energies of the atoms from H to Cl. Reference data are from [103]. We recently updated the basis set for this database to use a more complete basis set that includes core-polarization functions; in particular we now use the cc-pwCV5Z basis set [149] for H, He, and atoms from Be to Ne and from Al to Ar, while we used the cc-pCVQZ basis set [150] for Li, Be, Na and Mg atoms.

The solid-state physics energetic set, PE39, has two nonoverlapping primary subsets containing respectively cohesive energies and band gaps. For this set the lattice constants of the solids were reoptimized for each method studied. The primary subsets that compose the physics energetic database are:

- Solid-state cohesive energies (SSCE8): This set include the cohesive energies of eight solids: C, Si, SiC, Ge, NaCl, NaF, LiCl and LiF. We first used this database for the evaluation of the SOGGA functional [30]; reference data are taken from [104], and the geometry of each solid is optimized for each method. For this database we used the m-6-311G* basis set [151].

- Semiconductors band gaps (SBG31): This database was recently created for the evaluation of the performance of our M11-L functional [38]. It contains band gaps for four unary semiconductors from group 14 (C, Ge, Si, SiC), six binary semiconductors from groups 2 and 16 (MgS, MgSe, MgTe, BaS, BaSe, BaTe), fourteen binary semiconductors from groups 13 and 15 (BP, BAs, AlP, AlAs, AlSb, GaN, β-GaN, GaP, GaAs, GaSb, InN, InP, InAs, InSb), and seven binary semiconductors composed of a group 12 and an element from group 16 (ZnO, ZnS, ZnSe, ZnTe, CdS, CdSe, CdTe). Reference data are taken from [151-153]. In this database we follow the usual convention of comparing single-particle gaps from calculations (that is, band energy gaps, which are the crystal analogue of orbital energy gaps in molecules) to experimental optical gaps. Band gaps are calculated at the optimized geometry for each method. For this database we used the m-6-311G* basis set [151].

The chemistry structural database, CS20, is composed of two nonoverlapping primary subsets containing a total of 20 geometrical data. The primary structural databases are:

- Main-group hydrogenic bond lengths (MGHBL9): The MGHBL9 database of the structural set contains nine hydrogenic bond lengths [30]. For this database we used the 6-311+G(2df,2p) basis set [146].

- Main-group non-hydrogenic bond lengths (MGNHBL11): This set contains 11 non-hydrogenic bond lengths, and is composed by nine data from the older MGNHBL10 set [30] with the addition of the bond length of MgS [41]. For this database we used the 6-311+G(2df,2p) basis set [146].

Finally, the solid-state physics structural database, PS47, is composed of four nonoverlapping primary subsets containing a total of 47 structural data from 44 solids:

- Main group lattice constants (MLC4): The main group solid-state lattice constants set is composed of four main-group metals: Li, Na, K and Al. Reference data were taken from [104]. For this database we used the m-6-311G* basis set [151].

- Ionic lattice constants (ILC5): The ionic solid-state lattice constants set is composed of five ionic solids: NaCl, NaF, LiCl, LiF and MgO. Reference data were taken from [104]. For this database we used the m-6-311G* basis set [151].

- Transition metals lattice constants (TMLC4): The transition metals solid-state lattice constants set is composed of four transition metals: Cu, Rh, Pd and Ag. Reference data were taken from [104]. For this database we used the m-6-311G* basis set [151].

- Semiconductors lattice constants (SLC34): This database [38] is composed of the lattice constants of the same semiconductors in SBG31; the difference of three data from the band gaps database is explained by the fact that three solids with wurtzite structure (GaN, InN, ZnO) require the specification of two lattice constants. Reference data are equilibrium values [151,154-157] obtained by removing the zero-point anharmonic expansion, so that we can directly compare our calculated results with the experimental data.

## 4(*b*) Secondary and analytical databases

The secondary (level-3) subsets are described below:

- Atomization energies (AE6/11): This is a small subset of MGAE109/11 containing atomization energies of $SiH_4$, SiO, $S_2$, HCOCOH, propyne and cyclobutane.

- Small-B1 atomization energies (SB1AE97) and large-B1 atomization energies (LB1AE12): These databases are subsets of the MGAE109/11 database, and have been constructed according to the B1 diagnostic, which was originally developed [126] to give an indication of multireference [158] character. However we now recognize that it is a more general diagnostic signaling a "difficult case", perhaps because of multireference character but perhaps for other reasons. Nevertheless the B1 diagnostic as applied to MGAE109/11 probably does mainly differentiate single-reference and multireference cases, since the SB1AE97 set is composed of cases that are very likely single-reference, while LB1AE12 is composed of cases that can be multireference.

- Ionization potentials (IP13/03): The IP13/03 database [62,63,95,96] is composed of the 13 main group ionization potentials in IP21.

- Ionization potentials of metals (IPM8): The IPM8 database [70] is composed of eight ionization potentials of metal atoms and FeC, all from the IP21 database.

- Alkyl Bond Dissociation Energies (ABDE4/05): This database contains four bond dissociation energies of small R–X organic molecules, with R = methyl and isopropyl, and X = $CH_3$ and $OCH_3$. It is a subset of ABDE12.

- Larger set of Alkyl Bond Dissociation Energies of Molecules (ABDEL8): This set of alkyl bond dissociation energies [31] includes eight R–X bond dissociation energies of molecules with R = ethyl and tert-butyl and X = H, $CH_3$, $OCH_3$, OH. It is a subset of ABDE12.

- π-systems interaction energies (πIE3/06): The πIE3/06 database contains three isomeric energy differences between allene and propyne as well as higher homologs (which correspond to cumulenes and polyenes) [98,159].

- Proton affinities of conjugated polyenes (PA-CP5/06): The PA-CP5/06 database [98] contains proton affinities of five conjugated polyenes.

- Proton affinities of Schiff basis (PA-SB5/06): The PA-SB5/06 database [98] contains proton affinities of five conjugated Schiff bases.

- Heavy-atom transfer (HATBH12/08): The heavy-atom transfer database contains 12 reaction barrier heights involving heavy atoms.

- Nucleophilic substitution (NSBH16/08) The nucleophilic substitution database contains 16 barrier heights of nucleophilic substitution reactions.

- Unimolecular and association reactions (UABH10/08): The UAB10/08 database contains ten barrier heights of unimolecular and association reactions.

- Hydrogen bonds (HB6/04): The HB6/04 consists of binding energies of six hydrogen bonded dimers.

- Charge-transfer (CT7/04): The CT7/04 database consists of binding energies of seven charge transfer complexes.

- Dipole interactions (DI6/04): The DI6/04 database contains the binding energies of six dipole interaction complexes.

- Electrostatic dominated complexation energies (EDCE19): The EDCE19 database is a merger of the three previous secondary subsets of the noncovalent complexation energies database: HB6/04, CT7/04 and DI6/04. This secondary database was introduced in order to discriminate between complexation energies with different magnitudes within the NCCE31/05 database (e.g., π-π stacking and weak interaction complexation energies usually have magnitudes that are one order of magnitude smaller than the electrostatic dominated complexes).

- Weak interactions (WI7/05): The WI7/05 database [95] consists of the binding energies of seven weak interaction complexes, all of which are bound by dispersion-like interactions.

- π-π stacking (PPS5/05): The PPS5/05 database [95] consists of binding energies of five π-π stacking complexes.

The analytical (level-3) subsets are described below:

- Metal bond energies (MBE18): The MBE18 database comprehends all metal bond energies in our energetic broad chemistry set and is composed by all data in SRMBE13 and 5 data from MRBE10.

- Transition metal bond energies (TMBE15): The TMBE15 database collects the transition metals bond energies from the SRMBE13 (10 data) and MRBE10 (5 data). This database is directly derived from MBE18 by excluding three bond energies of molecules containing main-group metals.

- Diverse barrier heights (DBH24/08): The DBH24/08 subset [101] comes from a different subdivision within the barrier heights sets, and includes six hydrogen-transfer reactions from HTBH38/08, and 18 barrier heights from NHTBH38/08 representing six heavy-atom transfer, six nucleophilic substitution and six unimolecular and association reactions.

- Geometries of diatomic (DG6): The DG6 subset [41] comes from a different subdivision within the chemistry structural set, and includes six diatomic molecules: $H_2$, HF, OH, $N_2$, $Cl_2$, MgS. The first three molecules are from MGHBL9, while the last three molecules are from MGNHBL11. For this subset we used the 6-311+G(2df,2p) basis set [146].

- Solid-state lattice constants (SSLC18): The solid-state lattice constants comes from a different subdivision within the physics structural set, and includes a broad set of solids: four main-group metals (Li, Na, K, Al), five semiconductors (C, Si, SiC, Ge, GaAs), five ionic solids (NaCl, NaF, LiCl, LiF and MgO), and four transition metals (Cu, Rh, Pd, and Ag). For this subset we used the m-6-311G* basis set [151]. Note that SSLC18 and SLC34 are not completely nonoverlapping. In particular, five semiconductors occur in both SSLC18 and SLC34.

### 4(*c*) Statistical data

In general we report errors as a mean unsigned errors (MUEs), which are mean absolute deviations from the reference data. For five of the databases, in particular MGAE109/11, SB1AE97, LB1AE12, AE6, and DC9/12, we report errors as MUE per bond ($MUE_{PB}$). In these cases we first compute the MUE, and then we divide by the average number of bonds per datum (counting multiple bonds as well as single bonds as one). The average numbers are 4.71 for MGAE109/11, 5.10 for SB1AE97, 1.33 for LB1AE12, 4.67 for AE6, and 9.22 for DC9/12.

The CE345 database is of special interest because energetic properties have historically been the properties of most interest for density functional applications in chemistry. The MUE for this database is computed as follows:

$$
\begin{aligned}
\text{MUE(CE345)} = \{ &109 \times \text{MUE}_{PB}(\text{MGAE109/11}) + 13 \times \text{MUE(SRMBE13)} + \\
&10 \times \text{MUE(MRBE10)} + 6 \times \text{MUE(IsoL6/11)} + \\
&21 \times \text{MUE(IP21)} + 13 \times \text{MUE(EA13/03)} + 8 \times \text{MUE(PA8/06)} + \\
&12 \times \text{MUE(ABDE12)} + 7 \times \text{MUE(HC7/11)} + \\
&13 \times \text{MUE}(\pi\text{TC13}) + 38 \times \text{MUE(HTBH38/08)} + \\
&38 \times \text{MUE(NHTBH38/08)} + 31 \times \text{MUE(NCCE31/05)} + \\
&9 \times \text{MUE}_{PB}(\text{DC9/12}) + 17 \times \text{MUE(AE17)} \}/345
\end{aligned}
\tag{11}
$$

Two other average data are also used in the overall evaluation. The first is CExMR335, and is calculated as in equation (11) except excluding MRBE10. This set is useful for the evaluation of hybrid functionals, many of which should not be used for multireference systems, and the large error for these cases will dominate the average of BC345. The second is CExAE328, calculated including all subsets of CE345 except AE17. This set is useful for an alternative evaluation of those functionals that have large errors for the atomic energies (e.g., many GGA functionals such as PBE). The mean unsigned error for CExMR335 is calculated using the same formula and excluding MUE(MRBE10) (with 335 in the denominator), while that for CExAE328 is calculated excluding MUE(AE17) (with 328 in the denominator).

The mean unsigned error for PE39 is also calculated from the primary subsets as:

$$
\text{MUE(PE39)} = 1/39\{8*\text{MUE(SSCE8)} + 31*\text{MUE(SBG31)}\},
\tag{12}
$$

and the mean unsigned error for CS20 is calculated in a similar way as:

$$
\text{MUE(CS20)} = 1/20\{9*\text{MUE(MGHBL9)} + 11*\text{MUE(MGNHBL11)}\}.
\tag{13}
$$

The evaluation of functional performance using the physics structural set, PS47, are based on lattice constants, which by definition differ from the calculated nearest neighbor distances by a geometrical factor, and therefore they have different magnitude than the errors for bond lengths in molecules. If an approximate density functional were equally valid for solids and molecules, we would expect the MUEs for the lattice constants to be larger than those for molecular bond lengths by a factor of about 2.15, which is the average of the geometrical factors for the PS47 database. The mean errors for PS47 are calculated from the primary subsets as:

$$MUE(PS47) = 1/47\{4*MUE(MGLC4) + 5*MUE(ILC5) +$$
$$4*MUE(TMLC4) + 34*MUE(SLC34)\}. \tag{14}$$

## 5. Validation of DFT functionals

In this section we present results with a large number of density functionals, including all of our own second order functionals, the nonseparable gradient approximation, and the Minnesota functionals. This is primarily done to put the results of our most recent functionals in perspective, and to show strong and weak point of each method. It is worthwhile to point out at this point that, although we made use of parametrizations based on some of the databases presented above, none of our functionals has been parametrized using all the data in the four comprehensive databases. The functionals that we considered in this study are presented with the corresponding original reference or references in Table 2.

Table 2: Summary of the methods used for the calculations in this article (functionals are ordered in alphabetical order).

| Functional Name | Year | Reference(s) | Type$^a$ | $X^b$ |
|---|---|---|---|---|
| B1LYP | 1997 | [160] | GH-GGA | 25 |
| B3LYP | 1993 | [26,29,44] | GH-GGA | 20 |
| B3PW91 | 1992 | [26,43] | GH-GGA | 20 |
| B86LYP$^c$ | 1987 | [25,29] | GGA | 0 |
| B86P86$^c$ | 1986 | [25,161] | GGA | 0 |
| B86PW91$^c$ | 1991 | [18,25] | GGA | 0 |
| B97-3 | 2005 | [162] | GH-GGA | 26.93 |
| B97-D | 2006 | [137] | GGA+D | 0 |
| B98 | 1998 | [163] | GH-GGA | 21.98 |
| BB1K | 2004 | [164] | GH-mGGA | 42 |
| BLYP | 1988 | [26,29] | GGA | 0 |
| BMK | 2004 | [165] | mGGA | 42 |
| BP86 | 1988 | [26,161] | GGA | 0 |
| BPW91 | 1991 | [18,26] | GGA | 0 |
| CAM-B3LYP | 2004 | [53] | RSH-GGA | 19–65 |
| GVWN3 | 1980 | [2,15,17] | LSDA | 0 |
| GVWN5 | 1980 | [2,15,17] | LSDA | 0 |
| HCTH407 | 1997 | [166] | GGA | 0 |
| HF[d] | <1951[e] | [167][e] | WFT | 100 |
| HFLYP | 1987 | [29,167] | GH-GGA | 100 |
| HFPW91 | 1991 | [18,167] | GH-GGA | 100 |
| HSE06 | 2006 | [54,55] | RSH-GGA | 25–0 |
| LC-ωPBE | 2006 | [52] | RSH-GGA | 0–100 |
| M05 | 2005 | [62] | GH-mGGA | 28 |
| M05-2X | 2005 | [63] | GH-mGGA | 56 |
| M06 | 2008 | [65] | GH-mGGA | 27 |
| M06-2X | 2008 | [65] | GH-mGGA | 54 |
| M06-HF | 2006 | [66] | GH-mGGA | 100 |
| M06-L | 2006 | [64] | mGGA | 0 |
| M08-HX | 2008 | [67] | GH-mGGA | 52.23 |
| M08-SO | 2008 | [67] | GH-mGGA | 56.79 |
| M11 | 2011 | [68] | RSH-mGGA | 42.8–100 |
| M11-L | 2012 | [60] | mGGA | 0 |
| MN12-L | 2012 | [70] | mNGA | 0 |
| MN12-SX | 2012 | [56] | RSH-mGGA | 25–0 |
| MOHLYP | 2005 | [127] | GGA | 0 |
| MOHLYP2 | 2009 | [101] | GGA | 0 |
| MP2[d] | 1933 | [168] | WFT | 100 |
| MPW1B95 | 2004 | [169] | GH-mGGA | 31 |
| MPW1K | 2000 | [170] | GH-GGA | 42.8 |
| MPW1KCIS | 2005 | [100] | GH-mGGA | 15 |
| mPW1PW | 1997 | [171] | GH-GGA | 25 |
| MPW3LYP | 2004 | [169] | GH-GGA | 21.8 |
| MPWB1K | 2004 | [169] | GH-mGGA | 44 |
| MPWKCIS1K | 2005 | [100] | GH-mGGA | 41 |
| MPWLYP1M | 2005 | [127] | GH-GGA | 5 |
| MPWLYP1W | 2005 | [172] | GGA | 0 |
| mPWPW | 1997 | [171] | GGA | 0 |
| N12 | 2012 | [41] | NGA | 0 |
| N12-SX | 2012 | [56] | RSH-GGA | 25–0 |
| O3LYP | 2001 | [28,29] | GH-GGA | 20 |
| OLYP | 2001 | [28,29] | GGA | 0 |
| PBE | 1996 | [27] | GGA | 0 |
| PBE0[f] | 1996 | [173] | GH-GGA | 25 |
| PBE1KCIS | 2005 | [102] | GH-mGGA | 22 |
| PBE1W | 2005 | [172] | GGA | 0 |
| PBELYP1W | 2005 | [172] | GGA | 0 |
| PBEsol | 2008 | [33] | GGA | 0 |
| PW6B95 | 2005 | [95] | GH-mGGA | 28 |
| PW91 | 1991 | [18] | GGA | 0 |
| PWB6K | 2005 | [95] | GH-mGGA | 46 |
| revPBE | 1997 | [174] | GGA | 0 |
| revTPSS | 2009 | [175] | mGGA | 0 |
| RPBE | 1999 | [176] | GGA | 0 |
| SOGGA | 2008 | [30] | GGA | 0 |
| SOGGA11 | 2011 | [31] | GGA | 0 |
| SOGGA11-X | 2011 | [45] | GH-GGA | 40.15 |
| TPSS | 2002 | [177] | mGGA | 0 |
| TPSS1KCIS | 2005 | [71] | GH-mGGA | 13 |
| TPSSh | 2002 | [178] | GH-mGGA | 10 |
| TPSSLYP1W | 2005 | [172] | mGGA | 0 |
| VSXC$^g$ | 1998 | [179] | mGGA | 0 |
| τ-HCTH | 2002 | [180] | mGGA | 0 |
| τ-HCTHhyb | 2002 | [180] | GH-mGGA | 15 |
| ωB97 | 2008 | [69] | RSH-GGA | 0–100 |
| ωB97X | 2008 | [69] | RSH-GGA | 15.77–100 |
| ωB97X-D | 2008 | [181] | RSH-GGA+D | 22.2–100 |

[a]The acronyms in this column are: LSDA = local spin density approximation, GGA = generalized gradient approximations, +D = addition of molecular mechanic dispersion corrections, NGA = nonseparable gradient approximation, WFT = wave function theory, GH = global hybrid, RSH = range-separated hybrid (which can be either long-range-corrected or screened-exchange), mGGA = meta-GGA.

[b]$X$ denotes the percentage of HF exchange.

[c]The B86 exchange functional can be also called X$_{αβγ}$.

[d]Results for HF and MP2 wave function methods are also included in this evaluation, and are presented together with the global-hybrid GGA functionals.

[e]The HF method is named after the pioneering work by Hartree and Fock in the 1930s, however a unique reference for this method is not available, and we decided to report here the date and reference of the matrix HF formulation of Roothaan [167].

[f]The PBE0 functional can be also called PBE1PBE and PBEh, although PBEh is a deprecated name since it is also used for another functional.

[g]The VSXC functional can be also called VS98.

Although the present assessment aims for comprehensiveness, it is almost impossible to be truly complete. For example, not all functionals that have been proposed could be included. Functionals omitted include some less successful functionals, some promising functionals that we did not manage to get into our computer program yet with the time and resources available, some functionals omitted just to keep the tables compact enough to be readily comprehensible and the scope limited enough to allow reasonable discussion, and—as already mentioned—all functionals with nonlocal correlation simply because they raise so many new issues that they should have a separate assessment. Some examples of interesting unincluded functionals with local correlation are B86MGC [182], B97-1 [183], B3LYP* [184], LCgau-BOP [185], GauPBE [186], PW6B95-D3 [87], ωM05-D [187], and SXR12 [188]; and some examples of functionals with nonlocal correlation are given in section 2(d). Despite these limitations, the present comparison does involve a wide variety of functionals that illustrate most of the diverse features one may encounter.

<p align="center">5(<em>a</em>) <em>Results for CE345</em></p>

Results in terms of mean unsigned errors (MUEs) for the primary chemistry databases and their overall statistical data are presented in this section. For the purpose of an easier presentation, we grouped the functionals in eight classes, whose results are presented in different tables. The classes, with the respective tables are: LSDA and first generation GGA functionals (Table 3), where we define a first-generation GGA one that is published before 1993 (the year of publication of the B3LYP functional, which is a seam in the timeline of DFT); second generation GGA and NGA functionals (Table 4); first generation global hybrid GGA functionals (Table 5, once again first-generation functionals are functionals published before 1993); second generation global hybrid GGA functionals (Table 6); range-separated hybrid GGA functionals (Table 7); meta-GGA functionals (Table 8); first generation hybrid meta-GGA functionals (Table 9), where we define a first generation hybrid meta-GGA one that is published before the first Minnesota functional (M05); global and range-separated hybrid meta-GGA functionals published since M05 (Table 10). Within each table, functionals are ordered according to their year of publication.

Table 3: Mean unsigned errors (MUEs, in kcal/mol) for the primary chemistry databases in CE345 for LSDA and GGA functionals published before 1993 (year of publication of B3LYP).

| Functional: | GVWN5 | GVWN3 | B86P86 | B86LYP | BP86 | BLYP | B86PW91 | BPW91 | PW91 |
|---|---|---|---|---|---|---|---|---|---|
| Year of Publication: | 1980 | 1980 | 1986 | 1987 | 1988 | 1988 | 1991 | 1991 | 1991 |
| MGAE109/11 | 16.75 | 18.37 | 3.51 | 1.51 | 3.65 | 1.54 | 1.17 | 1.40 | 3.23 |
| SRMBE13 | 21.25 | 24.21 | 6.77 | 5.07 | 7.37 | 6.61 | 5.59 | 6.98 | 8.92 |
| MRBE10 | 30.31 | 32.82 | 11.78 | 12.21 | 13.55 | 6.53 | 16.47 | 12.67 | 14.78 |
| IsoL6/11 | 2.05 | 2.19 | 1.35 | 2.63 | 2.28 | 3.73 | 1.34 | 2.38 | 1.92 |
| IP21 | 9.89 | 20.26 | 9.12 | 5.91 | 8.67 | 6.56 | 6.78 | 6.23 | 7.39 |
| EA13/03 | 5.70 | 16.41 | 2.57 | 4.18 | 4.21 | 2.68 | 3.99 | 2.26 | 2.60 |
| PA8/06 | 5.07 | 4.55 | 3.59 | 2.68 | 1.41 | 1.58 | 4.89 | 1.88 | 1.30 |
| ABDE12 | 13.11 | 15.06 | 7.11 | 11.11 | 7.44 | 11.66 | 9.88 | 10.02 | 5.81 |
| HC7/11 | 21.45 | 23.50 | 6.33 | 19.48 | 9.95 | 27.39 | 5.02 | 10.77 | 4.55 |
| $\pi$TC13 | 4.80 | 4.66 | 10.36 | 10.24 | 5.85 | 6.07 | 11.65 | 7.08 | 5.73 |
| HTBH38/08 | 17.55 | 17.79 | 7.74 | 6.21 | 9.16 | 7.52 | 5.93 | 7.38 | 9.60 |
| NHTBH38/08 | 12.42 | 12.36 | 7.97 | 7.83 | 8.72 | 8.53 | 6.53 | 7.26 | 8.80 |
| NCCE31/05 | 3.17 | 3.31 | 1.35 | 1.26 | 1.46 | 1.55 | 1.20 | 1.69 | 1.37 |
| DC9/12 | 17.05 | 18.86 | 3.45 | 2.27 | 4.22 | 3.01 | 2.33 | 2.78 | 4.42 |
| AE17 | 421.13 | 309.99 | 29.64 | 24.85 | 16.92 | 8.68 | 24.67 | 11.95 | 4.63 |
| | | | | | | | | | |
| CE345 | 33.80 | 30.22 | 6.63 | 5.76 | 6.31 | 5.13 | 5.40 | 4.88 | 5.40 |
| CExMR335 | 33.90 | 30.15 | 6.48 | 5.57 | 6.09 | 5.09 | 5.07 | 4.65 | 5.12 |
| CExAE328 | 13.76 | 15.77 | 5.46 | 4.78 | 5.78 | 4.96 | 4.41 | 4.53 | 5.46 |

Table 4: Mean unsigned errors (MUEs, in kcal/mol) for the primary chemistry databases in CE345 for GGA and NGA functionals published after 1993 (year of publication of B3LYP).

| Functional: | PBE | HCTH407 | mPWPW | revPBE | RPBE | OLYP | MPWLYP1W | PBE1W | PBELYP1W | MOHLYP | B97-D | SOGGA | PBEsol | MOHLYP2 | SOGGA11 | N12 |
|---|---|---|---|---|---|---|---|---|---|---|---|---|---|---|---|---|
| Year of Publication: | 1996 | 1997 | 1997 | 1997 | 1999 | 2001 | 2005 | 2005 | 2005 | 2005 | 2006 | 2008 | 2008 | 2009 | 20011 | 2012 |
| MGAE109/11 | 3.07 | 1.12 | 2.05 | 1.68 | 1.99 | 0.91 | 1.34 | 2.46 | 1.70 | 2.51 | 0.85 | 7.82 | 7.94 | 17.86 | 1.68 | 1.27 |
| SRMBE13 | 6.91 | 6.45 | 6.26 | 5.81 | 5.90 | 9.29 | 8.43 | 8.72 | 7.58 | 9.06 | 24.21 | 10.81 | 10.93 | 17.57 | 12.31 | 8.97 |
| MRBE10 | 14.34 | 8.73 | 13.38 | 7.07 | 6.69 | 6.54 | 12.14 | 12.74 | 12.09 | 7.70 | 32.82 | 16.00 | 19.51 | 34.86 | 9.68 | 7.22 |
| IsoL6/11 | 1.98 | 3.02 | 2.16 | 2.82 | 2.99 | 3.44 | 3.67 | 2.42 | 3.95 | 4.31 | 1.73 | 1.89 | 1.55 | 6.80 | 1.73 | 1.73 |
| IP21 | 6.21 | 6.81 | 6.87 | 4.94 | 4.90 | 2.83 | 7.34 | 8.11 | 9.23 | 3.76 | 3.34 | 4.69 | 5.76 | 9.94 | 6.19 | 3.45 |
| EA13/03 | 2.27 | 3.70 | 2.30 | 2.39 | 2.37 | 3.60 | 3.20 | 3.67 | 5.39 | 3.67 | 2.46 | 2.70 | 2.16 | 13.20 | 5.23 | 4.21 |
| PA8/06 | 1.34 | 2.84 | 1.52 | 2.00 | 1.98 | 2.40 | 1.61 | 1.44 | 1.65 | 1.86 | 3.16 | 2.33 | 2.10 | 5.28 | 2.11 | 1.35 |
| ABDE12 | 6.14 | 10.75 | 8.11 | 12.16 | 12.73 | 11.51 | 10.42 | 8.04 | 10.49 | 14.86 | 8.64 | 4.28 | 3.47 | 32.71 | 6.93 | 5.63 |
| HC7/11 | 3.97 | 14.97 | 8.08 | 13.65 | 14.96 | 17.01 | 26.30 | 12.25 | 28.09 | 25.50 | 13.46 | 17.88 | 13.31 | 47.73 | 6.26 | 4.27 |
| πTC13 | 5.59 | 8.23 | 6.41 | 7.15 | 7.20 | 8.26 | 5.96 | 6.14 | 6.28 | 6.67 | 8.82 | 4.06 | 4.20 | 10.81 | 7.41 | 8.61 |
| HTBH38/08 | 9.31 | 5.48 | 8.43 | 6.58 | 6.43 | 5.63 | 8.17 | 8.45 | 7.85 | 5.48 | 7.17 | 12.88 | 12.69 | 4.14 | 6.57 | 6.94 |
| NHTBH38/08 | 8.42 | 6.29 | 8.03 | 6.82 | 6.82 | 5.26 | 9.05 | 8.20 | 8.79 | 5.80 | 6.34 | 9.68 | 9.86 | 3.41 | 4.32 | 6.86 |
| NCCE31/05 | 1.24 | 1.14 | 1.26 | 1.59 | 1.46 | 2.39 | 1.31 | 1.19 | 1.24 | 2.89 | 0.64 | 1.84 | 1.79 | 5.48 | 1.28 | 1.30 |
| DC9/12 | 4.27 | 3.85 | 3.17 | 3.80 | 4.16 | 3.62 | 2.33 | 3.36 | 2.54 | 6.44 | 2.85 | 9.24 | 8.93 | 22.27 | 3.33 | 3.02 |
| AE17 | 47.24 | 16.80 | 12.55 | 10.88 | 9.39 | 10.13 | 35.60 | 9.26 | 63.05 | 256.11 | 10.33 | 283.06 | 245.90 | 14.80 | 10.06 | 14.21 |
| | | | | | | | | | | | | | | | | |
| CE345 | 7.18 | 4.84 | 5.17 | 4.66 | 4.71 | 4.31 | 6.71 | 5.43 | 8.33 | 17.42 | 5.57 | 21.32 | 19.54 | 13.81 | 4.47 | 4.40 |
| CExMR335 | 6.97 | 4.72 | 4.92 | 4.59 | 4.65 | 4.25 | 6.55 | 5.21 | 8.22 | 17.71 | 4.75 | 21.48 | 19.54 | 13.18 | 4.32 | 4.32 |
| CExAE328 | 5.12 | 4.23 | 4.80 | 4.35 | 4.48 | 4.02 | 5.23 | 5.24 | 5.51 | 5.07 | 5.33 | 7.78 | 7.83 | 13.80 | 4.19 | 3.90 |

Table 5: Mean unsigned errors (MUEs, in kcal/mol) for the primary chemistry databases in CE345 for HF and MP2 and for global hybrid GGA functionals published before 1993 (year of publication of B3LYP).

| Functional:<br>Year of Publication: | HF<br><1951 | MP2<br>1933 | HFLYP<br>1987 | HFPW91<br>1991 | B3PW91<br>1992 | B3LYP<br>1993 |
|---|---|---|---|---|---|---|
| MGAE109/11 | 30.89 | 2.06 | 8.05 | 11.93 | 0.73 | 0.99 |
| SRMBE13 | 30.80 | 13.19 | 16.60 | 15.76 | 5.18 | 6.05 |
| MRBE10 | 132.37 | 115.51 | 98.28 | 116.66 | 24.17 | 21.68 |
| IsoL6/11 | 3.56 | 1.37 | 3.21 | 4.30 | 1.52 | 2.61 |
| IP21 | 19.69 | 9.69 | 9.24 | 26.25 | 4.87 | 5.71 |
| EA13/03 | 26.98 | 3.02 | 9.52 | 10.47 | 2.13 | 2.33 |
| PA8/06 | 3.26 | 0.93 | 3.32 | 8.41 | 1.89 | 1.02 |
| ABDE12 | 36.26 | 4.80 | 12.20 | 23.64 | 9.00 | 9.84 |
| HC7/11 | 16.04 | 5.68 | 16.49 | 25.72 | 4.35 | 16.80 |
| $\pi$TC13 | 9.79 | 2.24 | 9.65 | 16.74 | 7.00 | 6.03 |
| HTBH38/08 | 13.66 | 4.14 | 7.22 | 14.98 | 4.02 | 4.23 |
| NHTBH38/08 | 9.24 | 5.58 | 7.04 | 9.53 | 3.62 | 4.55 |
| NCCE31/05 | 2.26 | 0.53 | 0.64 | 2.13 | 1.15 | 0.96 |
| DC9/12 | 28.82 | 2.11 | 9.06 | 15.23 | 2.00 | 2.40 |
| AE17 | 191.60 | 134.86 | 8.53 | 10.25 | 4.83 | 18.29 |
| | | | | | | |
| CE345 | 31.98 | 13.43 | 10.48 | 15.78 | 3.47 | 4.60 |
| CExMR335 | 28.99 | 10.38 | 7.86 | 12.76 | 2.86 | 4.09 |
| CExAE328 | 23.78 | 7.16 | 10.61 | 16.11 | 3.41 | 3.90 |

Table 6: Mean unsigned errors (MUEs, in kcal/mol) for the primary chemistry databases in CE345 for global hybrid GGA functionals published after 1993 (year of publication of B3LYP).

| Functional: | PBE0 | mPW1PW | B1LYP | B98 | MPW1K | O3LYP | MPW3LYP | MPWLYP1M | B97-3 | SOGGA11-X |
|---|---|---|---|---|---|---|---|---|---|---|
| Year of Publication: | 1996 | 1997 | 1997 | 1998 | 2000 | 2001 | 2004 | 2005 | 2005 | 2011 |
| MGAE109/11 | 0.98 | 1.02 | 2.77 | 0.71 | 2.43 | 0.82 | 0.70 | 1.15 | 0.66 | 0.73 |
| SRMBE13 | 5.07 | 5.39 | 7.02 | 4.16 | 8.98 | 6.03 | 7.23 | 5.76 | 4.40 | 7.39 |
| MRBE10 | 27.87 | 30.16 | 37.62 | 18.04 | 51.59 | 16.70 | 21.94 | 6.99 | 23.76 | 35.12 |
| IsoL6/11 | 1.38 | 1.44 | 2.70 | 1.93 | 1.75 | 2.82 | 2.34 | 3.30 | 2.07 | 1.85 |
| IP21 | 3.37 | 3.93 | 3.35 | 3.56 | 3.54 | 3.38 | 5.07 | 6.22 | 3.47 | 3.49 |
| EA13/03 | 2.79 | 2.68 | 3.69 | 1.90 | 3.71 | 2.97 | 2.20 | 2.62 | 2.13 | 1.55 |
| PA8/06 | 1.19 | 1.77 | 1.06 | 1.53 | 2.41 | 2.36 | 0.97 | 1.47 | 2.54 | 1.85 |
| ABDE12 | 7.12 | 8.71 | 11.63 | 6.80 | 9.26 | 10.33 | 8.50 | 9.84 | 6.71 | 4.97 |
| HC7/11 | 9.40 | 6.70 | 17.64 | 8.46 | 12.98 | 12.52 | 14.00 | 22.96 | 7.58 | 7.27 |
| $\pi$TC13 | 6.11 | 6.85 | 6.03 | 7.13 | 7.28 | 8.00 | 5.46 | 5.60 | 7.37 | 6.08 |
| HTBH38/08 | 4.22 | 3.55 | 3.19 | 4.16 | 1.34 | 4.06 | 4.71 | 7.48 | 2.28 | 1.79 |
| NHTBH38/08 | 3.43 | 5.71 | 3.63 | 3.31 | 1.72 | 3.64 | 4.86 | 8.18 | 1.38 | 1.16 |
| NCCE31/05 | 0.69 | 0.73 | 0.97 | 0.70 | 0.58 | 1.93 | 0.84 | 1.19 | 0.98 | 0.63 |
| DC9/12 | 2.01 | 2.03 | 4.23 | 1.79 | 2.89 | 2.98 | 1.69 | 2.24 | 2.16 | 1.66 |
| AE17 | 38.57 | 10.78 | 9.88 | 4.91 | 9.93 | 5.57 | 5.19 | 12.09 | 6.80 | 4.98 |
| | | | | | | | | | | |
| CE345 | 5.20 | 4.17 | 5.07 | 3.11 | 4.86 | 3.68 | 3.80 | 4.86 | 3.00 | 3.09 |
| CExMR335 | 4.52 | 3.40 | 4.10 | 2.67 | 3.46 | 3.29 | 3.26 | 4.79 | 2.38 | 2.14 |
| CExAE328 | 3.48 | 3.84 | 4.84 | 3.03 | 4.61 | 3.59 | 3.74 | 4.50 | 2.81 | 3.00 |

Table 7: Mean unsigned errors (MUEs, in kcal/mol) for the primary chemistry databases in CE345 for range-separated hybrid GGA functionals.

| Functional: | CAM-B3LYP | LC-ωPBE | HSE06 | ωB97 | ωB97X | ωB97X-D | N12-SX |
|---|---|---|---|---|---|---|---|
| Year of Publication: | 2004 | 2006 | 2006 | 2008 | 2008 | 2008 | 2012 |
| MGAE109/11 | 0.70 | 0.89 | 0.88 | 0.59 | 0.51 | 0.52 | 0.76 |
| SRMBE13 | 7.81 | 9.65 | 5.17 | 8.65 | 8.13 | 5.71 | 6.25 |
| MRBE10 | 28.77 | 36.36 | 25.84 | 22.47 | 24.84 | 25.35 | 9.11 |
| IsoL6/11 | 2.07 | 1.90 | 1.25 | 1.48 | 1.55 | 1.15 | 1.78 |
| IP21 | 5.09 | 6.08 | 4.01 | 5.28 | 4.24 | 3.16 | 4.00 |
| EA13/03 | 2.06 | 2.15 | 2.77 | 2.58 | 2.01 | 1.86 | 2.99 |
| PA8/06 | 1.41 | 1.83 | 1.10 | 1.80 | 1.51 | 2.36 | 1.97 |
| ABDE12 | 6.80 | 6.54 | 7.74 | 3.85 | 4.45 | 4.52 | 5.30 |
| HC7/11 | 6.21 | 17.66 | 7.34 | 11.51 | 6.77 | 4.63 | 11.05 |
| πTC13 | 3.69 | 4.27 | 6.20 | 3.93 | 4.37 | 6.24 | 7.89 |
| HTBH38/08 | 3.18 | 1.18 | 4.23 | 1.88 | 2.01 | 2.36 | 3.71 |
| NHTBH38/08 | 2.61 | 2.36 | 3.73 | 2.41 | 2.89 | 3.74 | 2.83 |
| NCCE31/05 | 0.63 | 0.78 | 0.75 | 0.52 | 0.50 | 0.32 | 0.74 |
| DC9/12 | 1.13 | 1.48 | 1.96 | 1.38 | 1.06 | 1.14 | 1.19 |
| AE17 | 10.93 | 25.34 | 32.82 | 6.23 | 5.64 | 5.67 | 10.22 |
| | | | | | | | |
| CE345 | 3.57 | 4.72 | 4.88 | 3.03 | 2.93 | 2.95 | 3.20 |
| CExMR335 | 2.82 | 3.78 | 4.26 | 2.45 | 2.28 | 2.28 | 3.02 |
| CExAE328 | 3.20 | 3.66 | 3.44 | 2.87 | 2.80 | 2.82 | 2.84 |

Table 8: Mean unsigned errors (MUEs, in kcal/mol) for the primary chemistry databases in CE345 for meta-GGA and meta-NGA functionals

| Functional:<br>Year of Publication: | VSXC<br>1998 | τ-HCTH<br>2002 | TPSS<br>2002 | TPSSLYP1W<br>2005 | M06-L<br>2006 | revTPSS<br>2009 | M11-L<br>2012 | MN12-L<br>2012 |
|---|---|---|---|---|---|---|---|---|
| MGAE109/11 | 0.71 | 0.90 | 1.07 | 2.54 | 0.87 | 0.94 | 0.74 | 0.69 |
| SRMBE13 | 7.65 | 9.52 | 5.73 | 7.93 | 6.63 | 6.54 | 6.02 | 11.41 |
| MRBE10 | 8.77 | 10.60 | 6.79 | 6.62 | 8.19 | 7.54 | 6.70 | 7.35 |
| IsoL6/11 | 4.69 | 2.87 | 3.66 | 5.73 | 2.76 | 3.96 | 1.57 | 1.07 |
| IP21 | 3.79 | 4.55 | 4.14 | 5.96 | 3.89 | 3.92 | 4.60 | 3.51 |
| EA13/03 | 2.84 | 2.23 | 2.35 | 2.99 | 3.83 | 2.59 | 5.54 | 2.65 |
| PA8/06 | 2.02 | 3.19 | 2.66 | 2.54 | 1.88 | 2.79 | 2.17 | 1.91 |
| ABDE12 | 8.13 | 9.87 | 10.47 | 13.55 | 7.75 | 8.56 | 6.37 | 4.85 |
| HC7/11 | 14.63 | 14.32 | 10.48 | 30.14 | 3.35 | 6.42 | 2.42 | 2.58 |
| πTC13 | 8.37 | 8.76 | 8.12 | 8.34 | 6.69 | 7.85 | 5.14 | 5.32 |
| HTBH38/08 | 4.86 | 6.87 | 7.71 | 6.09 | 4.15 | 6.96 | 1.44 | 1.31 |
| NHTBH38/08 | 4.96 | 5.90 | 8.91 | 8.95 | 3.81 | 9.07 | 2.86 | 2.24 |
| NCCE31/05 | 2.22 | 1.11 | 1.17 | 1.23 | 0.58 | 1.14 | 0.56 | 0.46 |
| DC9/12 | 1.95 | 2.98 | 1.95 | 4.00 | 2.36 | 2.28 | 1.14 | 1.65 |
| AE17 | 49.90 | 17.14 | 18.04 | 86.16 | 7.04 | 23.81 | 21.81 | 9.73 |
| | | | | | | | | |
| CE345 | 5.92 | 4.83 | 4.97 | 9.44 | 3.16 | 5.06 | 3.32 | 2.61 |
| CExMR335 | 5.83 | 4.66 | 4.92 | 9.53 | 3.01 | 4.98 | 3.21 | 2.47 |
| CExAE328 | 3.65 | 4.20 | 4.31 | 5.48 | 2.97 | 4.10 | 2.36 | 2.25 |

Table 9: Mean unsigned errors (MUEs, in kcal/mol) for the primary chemistry databases in CE345 for hybrid meta-GGA functionals published before M05.

| Functional: | TPSSh | τ-HCTHhyb | BB1K | MPWB1K | MPW1B95 | BMK | TPSS1KCIS | MPWKCIS1K | MPW1KCIS | PBE1KCIS | PWB6K | PW6B95 |
|---|---|---|---|---|---|---|---|---|---|---|---|---|
| Year of Publication: | 2002 | 2002 | 2004 | 2004 | 2004 | 2004 | 2005 | 2005 | 2005 | 2005 | 2005 | 2005 |
| MGAE109/11 | 1.04 | 0.81 | 1.42 | 1.05 | 0.68 | 0.52 | 0.72 | 1.54 | 2.21 | 2.43 | 1.52 | 0.47 |
| SRMBE13 | 4.50 | 6.75 | 7.86 | 7.30 | 5.80 | 7.22 | 6.39 | 8.54 | 6.28 | 6.07 | 7.13 | 4.77 |
| MRBE10 | 16.22 | 8.60 | 44.19 | 45.22 | 29.56 | 31.53 | 15.18 | 46.16 | 14.08 | 21.63 | 47.77 | 26.71 |
| IsoL6/11 | 3.09 | 1.80 | 1.84 | 1.79 | 1.57 | 1.81 | 2.91 | 1.18 | 1.49 | 1.07 | 1.86 | 2.03 |
| IP21 | 3.36 | 4.03 | 2.62 | 3.29 | 1.93 | 4.09 | 3.01 | 2.73 | 4.27 | 3.29 | 2.92 | 3.22 |
| EA13/03 | 2.84 | 1.82 | 4.38 | 4.14 | 2.93 | 1.61 | 2.86 | 3.66 | 2.07 | 2.22 | 3.62 | 1.83 |
| PA8/06 | 2.76 | 1.91 | 1.45 | 1.17 | 1.03 | 1.05 | 2.26 | 1.40 | 1.37 | 1.30 | 1.23 | 1.16 |
| ABDE12 | 10.47 | 6.42 | 5.49 | 4.77 | 4.57 | 3.78 | 9.67 | 8.01 | 7.41 | 6.21 | 5.08 | 5.38 |
| HC7/11 | 6.89 | 6.89 | 9.52 | 11.21 | 6.57 | 6.05 | 8.48 | 9.72 | 5.71 | 4.30 | 9.74 | 4.15 |
| πTC13 | 8.06 | 7.45 | 5.99 | 5.68 | 5.38 | 4.58 | 7.49 | 6.11 | 5.55 | 5.03 | 5.85 | 5.82 |
| HTBH38/08 | 5.96 | 5.28 | 1.18 | 1.30 | 3.01 | 1.27 | 4.69 | 1.62 | 5.86 | 5.13 | 1.28 | 3.13 |
| NHTBH38/08 | 6.81 | 4.48 | 1.41 | 1.44 | 2.19 | 1.15 | 5.43 | 2.15 | 4.85 | 3.89 | 1.42 | 2.83 |
| NCCE31/05 | 0.94 | 0.85 | 0.84 | 0.36 | 0.49 | 0.95 | 0.88 | 0.65 | 0.93 | 0.69 | 0.27 | 0.50 |
| DC9/12 | 1.96 | 1.80 | 1.88 | 1.33 | 1.44 | 0.89 | 2.18 | 2.08 | 2.80 | 2.88 | 1.59 | 1.44 |
| AE17 | 15.26 | 6.03 | 15.48 | 15.99 | 16.76 | 16.73 | 20.99 | 5.03 | 6.13 | 31.36 | 65.50 | 98.53 |
| | | | | | | | | | | | | |
| CE345 | 4.50 | 3.29 | 4.20 | 4.10 | 3.54 | 3.34 | 4.38 | 3.99 | 3.96 | 5.13 | 6.69 | 7.51 |
| CExMR335 | 4.15 | 3.13 | 3.01 | 2.87 | 2.77 | 2.50 | 4.05 | 2.73 | 3.66 | 4.63 | 5.47 | 6.94 |
| CExAE328 | 3.95 | 3.15 | 3.63 | 3.49 | 2.87 | 2.65 | 3.53 | 3.94 | 3.86 | 3.78 | 3.66 | 2.81 |

Table 10: Mean unsigned errors (MUEs, in kcal/mol) for the primary chemistry databases in CE345 for global and range-separated hybrid meta-GGA functionals published after M05.

| Functional: | M05 | M05-2X | M06-HF | M06 | M06-2X | M08-HX | M08-SO | M11 | MN12-SX |
|---|---|---|---|---|---|---|---|---|---|
| Year of Publication: | 2005 | 2005 | 2006 | 2008 | 2008 | 2008 | 2008 | 2011 | 2012 |
| MGAE109/11 | 0.57 | 0.54 | 0.73 | 0.61 | 0.47 | 0.72 | 0.65 | 0.52 | 0.52 |
| SRMBE13 | 5.04 | 8.34 | 13.99 | 5.09 | 8.95 | 7.17 | 6.96 | 8.91 | 10.82 |
| MRBE10 | 14.91 | 38.98 | 66.62 | 18.44 | 40.70 | 46.62 | 44.95 | 41.38 | 10.51 |
| IsoL6/11 | 2.75 | 1.22 | 2.46 | 1.27 | 1.53 | 0.59 | 1.19 | 1.10 | 1.21 |
| IP21 | 5.15 | 4.43 | 7.43 | 4.05 | 2.82 | 4.42 | 3.56 | 8.50 | 5.22 |
| EA13/03 | 2.97 | 2.04 | 3.31 | 1.85 | 2.14 | 1.32 | 2.72 | 0.89 | 2.11 |
| PA8/06 | 2.27 | 1.43 | 2.28 | 1.84 | 1.65 | 1.08 | 1.64 | 1.03 | 1.16 |
| ABDE12 | 7.85 | 2.64 | 4.52 | 4.10 | 2.50 | 2.81 | 3.42 | 3.13 | 3.83 |
| HC7/11 | 7.71 | 3.64 | 2.29 | 2.78 | 2.15 | 4.89 | 4.60 | 3.74 | 2.21 |
| $\pi$TC13 | 5.69 | 3.06 | 2.05 | 4.40 | 1.49 | 1.87 | 1.84 | 2.24 | 3.24 |
| HTBH38/08 | 1.94 | 1.35 | 2.07 | 1.98 | 1.14 | 0.72 | 1.07 | 1.30 | 0.95 |
| NHTBH38/08 | 2.07 | 1.81 | 2.53 | 2.33 | 1.22 | 1.22 | 1.23 | 1.28 | 1.35 |
| NCCE31/05 | 0.49 | 0.28 | 0.41 | 0.41 | 0.29 | 0.35 | 0.37 | 0.26 | 0.30 |
| DC9/12 | 1.81 | 0.99 | 1.38 | 1.01 | 1.03 | 0.95 | 1.08 | 0.80 | 1.20 |
| AE17 | 10.65 | 10.09 | 12.42 | 4.45 | 2.14 | 4.10 | 3.76 | 8.88 | 4.52 |
| | | | | | | | | | |
| CE345 | 3.03 | 3.19 | 4.83 | 2.42 | 2.59 | 2.94 | 2.93 | 3.33 | 2.16 |
| CExMR335 | 2.68 | 2.13 | 2.99 | 1.94 | 1.45 | 1.64 | 1.67 | 2.19 | 1.91 |
| CExAE328 | 2.64 | 2.85 | 4.45 | 2.32 | 2.62 | 2.89 | 2.89 | 3.05 | 2.04 |

Let's first analyze the performances for the most basic databases that are of most general interest to chemists, in particular those for atomization energies, barrier heights, and noncovalent interactions. The best functionals for the MGAE109/11 database are in general hybrid meta-GGAs with a high percentage of HF exchange; the most successful functionals are M06-2X, PW6B95, $\omega$B97X, MN12-SX, $\omega$B97X-D, M11 and BMK, all with average errors in the range 0.50±0.03 kcal/mol per bond. Local functionals at the LSDA and GGA level are on average not successful for atomization energies, with mean errors higher than 1 kcal/mol per bond. Some meta-GGA and meta-NGA functionals, however, are capable of providing acceptable performance even without nonlocal HF exchange, in particular MN12-L, VSXC and M11-L have errors below 0.75 kcal/mol per bond, which is comparable to many hybrid functionals.

The situation for the barrier heights databases (HTBH38/08 and NHTBH38/08) is similar in some respects to that for atomization energies, but with some key differences, since barrier heights seems to be sensitive not only to the percentage of HF exchange, but also to the quality of the density functional [101]. The most successful functionals for this case are again the hybrid meta-GGA functionals with a high percentage of HF exchange. Among local functionals, MN12-L and M11-L stand out with mean errors close to those of hybrid functionals (< 3 kcal/mol), while all other local functionals, including M06-L, which at one point in time was the best local functional for barrier heights, are above 4 kcal/mol.

Next consider the noncovalent interactions. For the NCCE31/05 database it is interesting to notice how important is the use of the kinetic energy density. Meta-GGA functionals are on average much more successful than GGA and LSDA, clearly showing that the kinetic energy density is a crucial ingredient for this property. M11, PWB6K, M05-2X, M06-2X, and MN12-SX all stand out with MUEs of 0.30 kcal/mol or less.

As far as the other molecular properties, we notice that, as expected, local functionals are much more successful than hybrid functionals for the multireference database, for which hybrid functionals with a high percentage of HF exchange fail badly. The rearrangement of our databases shows more clearly than previous analyses which functionals represent progress in treating multireference systems. It is clear when considering all databases in CE345 that the most successful functionals considering all databases in CE345 are range-separated hybrid functionals (ωB97X), local meta-GGA and meta-NGA functionals (M06-L, M11-L, and MN12-L), and hybrid meta-GGA functionals with a moderate percentage of HF exchange (MN12-SX, M05, M06).

Atomic energies, as represented by the AE17 database, deserve special consideration. Some exchange functionals, such as the B88 and OptX functionals were optimized to fit HF exchange energies. Since the exchange energy is much larger in magnitude than the correlation energy, this helps to get better results for the atomic energies, but it is not sufficient for getting the best results for at least two reasons: (i) The HF exchange energy is "exact" for HF wave functions but not for correlated ones, and (ii) one must also have an accurate correlation potential. Nevertheless, the OptX functional does give relatively accurate atomic energies [5]. The tables show MUEs in atomic energies of 8.67 and 10.13 kcal/mol, respectively, for BLYP and OLYP, and several other functionals do better than this. Functionals with MUEs below 5.2 kcal/mol for AE17 are PW91, B3PW91, B98, MPW3LYP, SOGGA11-X, MPWKCIS1K, M06-2X, M08-HX, M08-SO, and MN12-SX. Based on the GMTKN30 database, Goerigk and Grimme [87] found "no statistical correlation between a functional's accuracy for atomization energies and the performance for chemically more relevant reaction energies." Nevertheless, if one wants to obtain the right answer for the right reason, without relying on cancellation of errors, obtaining a reasonable value for the atomic energy along with good properties for molecules and solids is a worthwhile goal, and it may be helpful when one considers even broader databases. For those who wish to evaluate the performance of density functionals without considering absolute atomic energies, the tables also provide the MUE for the CExAE328 database, in which atomic energies are excluded.

*5(b) Results for PE39*

Results for the solid-state physics energetic set and its primary databases are presented in this section. Because of the high expense of long-range HF exchange, global hybrid and long-range-corrected hybrid functionals are not included in this evaluation. Results for the physics set are presented for LSDA, GGA, NGA, screened-exchange GGA, and meta-GGA functionals in Table 11.

Results for the cohesive energies database, SSCE8, show that LSDA has the worst performance for solid cohesive energies, while GGA-type functionals that are especially accurate for lattice constant calculations (such as SOGGA and PBEsol) are not high performers for solid-state cohesive energy. On the other side, some functionals that work better for chemical energetics (e.g. SOGGA11, PBE, N12-SX, and MN12-L) are successful for the cohesive energies database.

Understanding the use of DFT for band gap prediction has a long history [189] and is of great practical importance [38,151,190,191]. The incorrect prediction of band gaps in solids is an especially serious problem for treating defects, where various corrections are introduced, but they lead to inconsistent results for the positions of defect levels relative to band edges [192]. These problems are minimized, even if they do not go away, by using a density functional that gives a smaller error in the band gap. Our results for the band gap database show that almost all local functionals have MUEs in band gaps of $1.0 \pm 0.2$ eV. Notable exceptions are the two Minnesota meta-GGA and meta-NGA functionals, M06-L and M11-L, which are more precise than the other functionals. The hybrid HSE06 and N12-SX functionals are the best performers for this database, although M11-L provides very good performance for a local functional.

It is interesting that TPSSLYP1W performs better than revTPSS for two of the three databases in table 11, especially when one considers that the LYP correlation functional does not satisfy the UEG constraint as the density inhomogeneity vanishes and would not have been expected to be compatible with TPSS exchange. The single parameter in TPSSLYP1W was fit to water clusters.

The best performers for PE39 are N12-SX, HSE06, MN12-SX, M11-L, M06-L, and MN12-L.

Table 11: Mean unsigned errors (MUEs) for the physics energetic database, PE39, and its two subdatabases. (All values in eV; functionals are ordered according to increasing MUE for PE39).

|  | SSCE8 | SBG31 | PE39 |
|---|---|---|---|
| N12-SX | 0.11 | 0.26 | 0.23 |
| HSE06 | 0.11 | 0.26 | 0.23 |
| MN12-SX | 0.15 | 0.32 | 0.29 |
| M11-L | 0.24 | 0.54 | 0.48 |
| M06-L | 0.17 | 0.73 | 0.62 |
| MN12-L | 0.11 | 0.84 | 0.69 |
| TPSS | 0.22 | 0.85 | 0.72 |
| SOGGA11 | 0.07 | 0.89 | 0.72 |
| HCTH407 | 0.24 | 0.89 | 0.76 |
| MOHLYP2 | 0.33 | 0.88 | 0.77 |
| τ-HCTH | 0.22 | 0.91 | 0.77 |
| TPSSLYP1W | 0.32 | 0.90 | 0.78 |
| OLYP | 0.36 | 0.90 | 0.79 |
| PBE | 0.11 | 0.98 | 0.80 |
| N12 | 0.13 | 0.99 | 0.81 |
| revTPSS | 0.13 | 1.00 | 0.82 |
| MOHLYP | 0.27 | 1.02 | 0.87 |
| PW91 | 0.10 | 1.11 | 0.90 |
| mPWPW | 0.10 | 1.11 | 0.90 |
| B86PW91 | 0.16 | 1.10 | 0.91 |
| RPBE | 0.28 | 1.07 | 0.91 |
| revPBE | 0.27 | 1.08 | 0.91 |
| BP86 | 0.12 | 1.12 | 0.91 |
| BPW91 | 0.20 | 1.10 | 0.92 |
| VSXC | 0.52 | 1.03 | 0.93 |
| B86P86 | 0.22 | 1.12 | 0.94 |
| SOGGA | 0.27 | 1.14 | 0.96 |
| B86LYP | 0.26 | 1.15 | 0.97 |
| PBEsol | 0.31 | 1.14 | 0.97 |
| PBELYP1W | 0.48 | 1.10 | 0.97 |
| BLYP | 0.37 | 1.14 | 0.98 |
| MPWLYP1W | 0.40 | 1.14 | 0.99 |
| PBE1W | 0.43 | 1.15 | 1.00 |
| GVWN5 | 0.70 | 1.14 | 1.05 |
| GVWN3 | 0.70 | 1.14 | 1.05 |

*5(c) Results for CS20*

Results for the chemistry structural set and its primary databases are presented in Table 12.

Table 12: Mean unsigned errors (MUEs) for the primary chemistry structure databases in CS20 and its subdatabases. (All values in Å; functionals are ordered according to increasing MUE for CS20).

| Functional: | MGHBL9 | MGNHBL11 | CS20 | Functional: | MGHBL9 | MGNHBL11 | CS20 |
|---|---|---|---|---|---|---|---|
| VSXC | 0.002 | 0.004 | 0.003 | N12 | 0.006 | 0.011 | 0.008 |
| M06-L | 0.002 | 0.006 | 0.004 | M08-SO | 0.006 | 0.011 | 0.008 |
| TPSS1KCIS | 0.003 | 0.005 | 0.004 | PW91 | 0.010 | 0.007 | 0.008 |
| HCTH407 | 0.003 | 0.005 | 0.004 | M08-HX | 0.003 | 0.014 | 0.008 |
| O3LYP | 0.003 | 0.006 | 0.004 | B86LYP | 0.006 | 0.011 | 0.009 |
| TPSSh | 0.004 | 0.005 | 0.004 | M05-2X | 0.002 | 0.015 | 0.009 |
| τ-HCTHhyb | 0.003 | 0.006 | 0.005 | M11 | 0.004 | 0.013 | 0.009 |
| τ-HCTH | 0.004 | 0.006 | 0.005 | B86P86 | 0.005 | 0.012 | 0.009 |
| MPW1KCIS | 0.004 | 0.006 | 0.005 | mPWPW | 0.010 | 0.008 | 0.009 |
| PBE1KCIS | 0.002 | 0.007 | 0.005 | BPW91 | 0.010 | 0.009 | 0.009 |
| B98 | 0.002 | 0.009 | 0.005 | PBEsol | 0.014 | 0.006 | 0.010 |
| B3LYP | 0.002 | 0.009 | 0.006 | PBE | 0.011 | 0.008 | 0.010 |
| B97-3 | 0.001 | 0.010 | 0.006 | GVWN5 | 0.012 | 0.008 | 0.010 |
| ωB97X-D | 0.002 | 0.009 | 0.006 | GVWN3 | 0.013 | 0.006 | 0.010 |
| mPW1PW | 0.001 | 0.010 | 0.006 | SOGGA | 0.014 | 0.005 | 0.010 |
| SOGGA11 | 0.004 | 0.008 | 0.006 | PBE1W | 0.010 | 0.010 | 0.010 |
| PBE0 | 0.002 | 0.010 | 0.006 | MPWLYP1M | 0.008 | 0.011 | 0.010 |
| B1LYP | 0.002 | 0.010 | 0.006 | LC-ωPBE | 0.003 | 0.017 | 0.010 |
| MPW3LYP | 0.002 | 0.010 | 0.006 | M11-L | 0.008 | 0.012 | 0.010 |
| HSE06 | 0.002 | 0.009 | 0.006 | B86PW91 | 0.007 | 0.014 | 0.010 |
| MN12-SX | 0.004 | 0.011 | 0.007 | BP86 | 0.012 | 0.010 | 0.011 |
| ωB97X | 0.002 | 0.011 | 0.007 | TPSSLYP1W | 0.007 | 0.014 | 0.011 |
| SOGGA11-X | 0.002 | 0.012 | 0.007 | MPWKCIS1K | 0.005 | 0.019 | 0.012 |
| M06-2X | 0.002 | 0.012 | 0.007 | revPBE | 0.012 | 0.013 | 0.012 |
| MP2 | 0.002 | 0.012 | 0.007 | RPBE | 0.012 | 0.014 | 0.013 |
| CAM-B3LYP | 0.002 | 0.012 | 0.007 | MPWLYP1W | 0.010 | 0.016 | 0.013 |
| PW6B95 | 0.002 | 0.011 | 0.007 | PBELYP1W | 0.009 | 0.016 | 0.013 |
| M05 | 0.002 | 0.012 | 0.007 | MPW1K | 0.006 | 0.021 | 0.013 |
| TPSS | 0.007 | 0.007 | 0.007 | BB1K | 0.006 | 0.020 | 0.013 |
| ωB97 | 0.002 | 0.012 | 0.007 | M06-HF | 0.003 | 0.023 | 0.013 |
| OLYP | 0.007 | 0.008 | 0.007 | BLYP | 0.011 | 0.016 | 0.013 |
| M06 | 0.003 | 0.012 | 0.007 | MPWB1K | 0.006 | 0.021 | 0.014 |
| B3PW91 | 0.003 | 0.012 | 0.007 | PWB6K | 0.008 | 0.022 | 0.015 |
| revTPSS | 0.009 | 0.007 | 0.008 | MOHLYP | 0.017 | 0.020 | 0.019 |
| B97-D | 0.007 | 0.008 | 0.008 | MOHLYP2 | 0.013 | 0.025 | 0.019 |
| BMK | 0.002 | 0.013 | 0.008 | HF | 0.013 | 0.030 | 0.021 |
| MPW1B95 | 0.002 | 0.013 | 0.008 | HFLYP | 0.022 | 0.045 | 0.033 |
| MN12-L | 0.003 | 0.012 | 0.008 | HFPW91 | 0.022 | 0.049 | 0.036 |
| N12-SX | 0.004 | 0.013 | 0.008 | | | | |

Results for the chemistry structural database show that the most successful functionals are either local or hybrids with a moderate percentage of HF exchange. However, many functionals perform satisfactorily, with 59 functionals having mean unsigned errors of 0.010 Å or less.

5(*d*) *Results for PS47*

Results for the solid-state physics structural set and its primary databases are presented in this section. As for the PE39 database, global hybrid and long-range-corrected hybrid functionals are not included in this evaluation, because evaluating the long-range HF exchange is expensive and not practical. Results are presented in Table 13.

Table 13: Mean unsigned errors (MUEs) for the primary physics structure databases in PS47 and its subdatabases. (All values in Å, functionals are ordered according to increasing MUE for PS47).

|  | MGLC4 | ILC5 | TLC4 | SLC34 | PS47 |
|---|---|---|---|---|---|
| SOGGA | 0.023 | 0.020 | 0.020 | 0.023 | 0.022 |
| PBEsol | 0.023 | 0.027 | 0.019 | 0.031 | 0.029 |
| N12-SX | 0.035 | 0.020 | 0.019 | 0.034 | 0.031 |
| N12 | 0.019 | 0.020 | 0.030 | 0.035 | 0.032 |
| MN12-L | 0.024 | 0.025 | 0.011 | 0.039 | 0.034 |
| HSE | 0.053 | 0.026 | 0.048 | 0.040 | 0.040 |
| MN12-SX | 0.050 | 0.031 | 0.028 | 0.044 | 0.042 |
| revTPSS | 0.014 | 0.065 | 0.018 | 0.048 | 0.044 |
| VSXC | 0.019 | 0.020 | 0.030 | 0.056 | 0.047 |
| GVWN5 | 0.090 | 0.084 | 0.040 | 0.045 | 0.053 |
| GVWN3 | 0.090 | 0.084 | 0.040 | 0.045 | 0.053 |
| TPSS | 0.053 | 0.068 | 0.027 | 0.070 | 0.065 |
| M11-L | 0.043 | 0.054 | 0.077 | 0.069 | 0.066 |
| M06-L | 0.142 | 0.050 | 0.056 | 0.077 | 0.078 |
| PBE | 0.034 | 0.085 | 0.064 | 0.084 | 0.078 |
| PW91 | 0.043 | 0.073 | 0.048 | 0.094 | 0.083 |
| mPWPW | 0.058 | 0.091 | 0.051 | 0.101 | 0.092 |
| BPW91 | 0.073 | 0.106 | 0.065 | 0.100 | 0.095 |
| BP86 | 0.046 | 0.086 | 0.063 | 0.107 | 0.096 |
| B86P86 | 0.044 | 0.099 | 0.086 | 0.114 | 0.104 |
| B86PW91 | 0.060 | 0.117 | 0.088 | 0.117 | 0.109 |
| τ-HCTH | 0.133 | 0.168 | 0.041 | 0.112 | 0.114 |
| SOGGA11 | 0.221 | 0.206 | 0.099 | 0.092 | 0.116 |
| PBE1W | 0.111 | 0.150 | 0.100 | 0.126 | 0.125 |
| revPBE | 0.103 | 0.153 | 0.069 | 0.140 | 0.132 |
| TPSSLYP1W | 0.101 | 0.123 | 0.111 | 0.150 | 0.140 |
| OLYP | 0.124 | 0.197 | 0.071 | 0.149 | 0.145 |
| RPBE | 0.109 | 0.166 | 0.077 | 0.155 | 0.146 |
| HCTH407 | 0.115 | 0.206 | 0.065 | 0.160 | 0.153 |

| | | | | |
|---|---|---|---|---|
| PBELYP1W | 0.121 | 0.165 | 0.080 | 0.164 | 0.153 |
| MPWLYP1W | 0.130 | 0.201 | 0.080 | 0.159 | 0.154 |
| BLYP | 0.072 | 0.116 | 0.121 | 0.178 | 0.158 |
| B86LYP | 0.080 | 0.128 | 0.145 | 0.183 | 0.165 |
| MOHLYP | 0.148 | 0.198 | 0.116 | 0.192 | 0.182 |
| MOHLYP2 | 0.220 | 0.202 | 0.189 | 0.254 | 0.240 |

The solid-state physics structural database is very interesting because a great amount of recent attention has been devoted to this problem in the physics community. Results in Table 13 show that functionals that SOGGA, PBEsol, and HSE06 are among the top performers for this database, ranking higher than in other databases. However, the performance of the N12 functional is noteworthy because not only is it the third best performer for this database, but also it is the top performer for CE345 among functionals that depend only on the density and its gradient. N12-SX and MN12-L are also noteworthy for having outstanding performance on both CE345 and PS47. Among GGA functionals SOGGA11 is by far the best functional for CE345, however its performance for the solid-state structural database is disappointing. This observation proved wrong the earlier hypothesis that functionals that are correct to second order will produce good solid-state lattice constants, and this observation is the main finding that brought us to the development of the N12 functional. This situation indicates a single GGA functional is too limited to provide good accuracy for both chemistry and solid-state physics, and only a more flexible functional form was able to overcome the problem. At the meta-GGA level, MN12-L provides optimal performance for PS47, followed by revTPSS, but the performance of the latter for the chemistry database is a little disappointing (although not terrible compared to other local functionals). The opposite is true for our early Minnesota functionals: M06-L and M11-L are top performers for the chemistry database, but their accuracies for PS47 are a little disappointing (although not as terrible as many other functionals that perform well for chemistry). MN12-L though is outstanding for both chemistry and physics.

## 5(*e*) Secondary and analytical databases

The analysis of the results for secondary and analytical databases shows other interesting patterns. The main characteristics of this analysis are reported below, while detailed results are presented in the supporting information.

One of the most interesting assessments in light of the current heavy activity by several groups in adding molecular mechanics to density functional theory is the area of attractive noncovalent interactions. Results for the secondary subsets of the noncovalent interactions database are reported in Table 14 for the 25 functionals that perform best on NCCE31/05 along with—for comparison—the MP2 WFT method and the popular PBE and B3LYP functionals.

From the data we notice immediately that the first 19 functionals have either range-separation in the exchange or a meta-GGA functional form, or both, clearly demonstrating the desirability of going beyond the incomplete description at the GGA and global hybrid GGA levels, which have been incapable of providing a quantitative treatment of energetic properties with a strong dependence on medium-range correlation energy. The first five functionals (M11, PWB6K, M05-2X, M06-2X, and MN12-SX) show excellent performance without molecular mechanics, and have MUEs of 0.45 kcal/mol or less for all six of the secondary noncovalent complexation energy databases. The two best global hybrid GGAs for noncovalent interactions are SOGGA11-X and MPW1K.

A clear trend can be seen in the results for noncovalent interactions for local functionals and functionals with a low percentage of HF exchange, in that they perform worse for the charge-transfer database (CT7/04), as expected by their local (or very small nonlocal) character. However, the local MN12-L is better for CT7/04 than MP2 or HFLYP, which both have 100% HF exchange or than B3LYP, which has 20%. Performance for the other secondary noncovalent complexation energy databases seems more influenced by the quality of the density functional form than simply by the percentage of nonlocal exchange. For the weak interactions (WI7/05) and the $\pi$-$\pi$ stacking (PPS5/05) databases it is interesting to note the performance of MP2, which is the best method for weak interactions, but the worst for $\pi$-$\pi$ stacking. Another interesting feature of Table 14, in light of the frequently heard (but erroneous) comment that one can only obtain good results for noncovalent interactions by heavy parametrization, is the good performance of the lightly parametrized PWB6K, MPW1B95, and PW6B95 functionals.

Table 14: Mean unsigned errors (MUEs) for the noncovalent complexation energies database (NCCE31/05) and its secondary databases for the 25 best functionals on NCCE31/05 plus MP2, B3LYP, and PBE. (All values in kcal/mol; functionals are ordered according to2increasing MUE for NCCE31/05).

| | NCCE31/05 | HB6/04 | CT7/04 | DI6/04 | EDCE19 | WI7/05 | PPS5/05 |
|---|---|---|---|---|---|---|---|
| M11 | 0.26 | 0.37 | 0.30 | 0.33 | 0.33 | 0.09 | 0.22 |
| PWB6K | 0.27 | 0.44 | 0.26 | 0.24 | 0.31 | 0.18 | 0.23 |
| M05-2X | 0.28 | 0.39 | 0.45 | 0.23 | 0.36 | 0.10 | 0.22 |
| M06-2X | 0.29 | 0.43 | 0.37 | 0.22 | 0.34 | 0.18 | 0.21 |
| MN12-SX | 0.30 | 0.44 | 0.40 | 0.34 | 0.40 | 0.13 | 0.19 |
| ωB97X-D | 0.32 | 0.44 | 0.28 | 0.27 | 0.33 | 0.04 | 0.69 |
| M08-HX | 0.35 | 0.47 | 0.59 | 0.25 | 0.44 | 0.09 | 0.36 |
| MPWB1K | 0.36 | 0.41 | 0.23 | 0.52 | 0.38 | 0.08 | 0.66 |
| M08-SO | 0.37 | 0.46 | 0.66 | 0.29 | 0.48 | 0.09 | 0.30 |
| M06 | 0.41 | 0.26 | 1.07 | 0.27 | 0.56 | 0.18 | 0.18 |
| M06-HF | 0.41 | 0.63 | 0.35 | 0.53 | 0.50 | 0.22 | 0.38 |
| MN12-L | 0.46 | 0.73 | 0.70 | 0.42 | 0.62 | 0.18 | 0.27 |
| M05 | 0.49 | 0.57 | 0.66 | 0.23 | 0.50 | 0.14 | 0.98 |
| MPW1B95 | 0.49 | 0.50 | 0.48 | 0.50 | 0.49 | 0.09 | 1.06 |
| ωB97X | 0.50 | 0.94 | 0.67 | 0.58 | 0.73 | 0.03 | 0.29 |
| PW6B95 | 0.50 | 0.53 | 0.70 | 0.40 | 0.55 | 0.14 | 0.82 |
| ωB97 | 0.52 | 1.02 | 0.56 | 0.64 | 0.73 | 0.07 | 0.35 |
| MP2 | 0.53 | 0.26 | 0.73 | 0.45 | 0.50 | 0.07 | 1.27 |
| M11-L | 0.56 | 0.78 | 0.94 | 0.30 | 0.69 | 0.31 | 0.43 |
| M06-L | 0.58 | 0.22 | 1.78 | 0.32 | 0.83 | 0.17 | 0.21 |
| MPW1K | 0.58 | 0.33 | 0.44 | 0.52 | 0.43 | 0.20 | 1.70 |
| SOGGA11-X | 0.63 | 0.24 | 0.21 | 0.54 | 0.32 | 0.42 | 2.07 |
| CAM-B3LYP | 0.63 | 0.56 | 0.48 | 0.40 | 0.48 | 0.14 | 1.87 |
| HFLYP | 0.64 | 1.50 | 0.87 | 0.22 | 0.87 | 0.15 | 0.47 |
| B97-D | 0.64 | 0.40 | 1.89 | 0.35 | 0.93 | 0.15 | 0.23 |
| MPWKCIS1K | 0.65 | 0.50 | 0.48 | 0.60 | 0.52 | 0.18 | 1.81 |
| B3LYP | 0.96 | 0.60 | 0.71 | 0.78 | 0.70 | 0.30 | 2.89 |
| PBE | 1.24 | 0.45 | 2.97 | 0.45 | 1.38 | 0.14 | 2.23 |

Another interesting set of databases to compare is the group of two alkyl bond dissociation databases, the weak interaction database, and the π–π stacking database. These databases all depend on having a good treatment of medium-range correlation energy, but in different ways. Performing better than average on all four databases is a difficult challenge. This challenge is met by only three local functionals (namely SOGGA, M11-L, and MN12-L), although two others (PW91 and M06-L) just miss this mark, and by only two global-hybrid GGAs (PBE0 and B98). However, this test is passed by all the range-separated functionals, and it is passed by nine hybrid meta-GGAs (PW6B95, M05-2X, M06-2X, M06, M06-2X, M08-HX, M08-SO, and MN12-SX ).

Another set of related databases is LB1AE12, DC9, TMBE15, and MRBE10. Only a functional that can handle multireference systems can do well on these databases, and we ask how many functionals do better than average on LB1AE12 and DC9 and also have an MUE below 10 kcal/mol on both TMBE15 and MRBE10. Only eight functionals succeed, seven local functionals (OLYP, RPBE, revPBE, N12, M06-L, M11-L, and MN12-L) and one nonlocal (N12-SX). Notice that this set has no overlap with the successful performers in the previous paragraph, which provides another example of the difficulty of finding a universal functional. As mentioned above, the metal databases are still rather small, and further work is needed to draw more definitive conclusions about performance for metal-containing molecules.

Ionization potentials and proton affinities both require good accuracy for cations. We then ask which functionals have an MUE below 5 kcal/mol for IP21 and for both proton affinity databases. Surprisingly, only three local functionals (SOGGA11, M11-L, and MN12-L) and four hybrid functionals (M05-2X, M06-2X, M08-HX, and M08-SO) pass this test—with M06-2X getting all three MUEs below 3 kcal/mol.

The extra databases allow us to ask which density functionals perform well for all four kinds of barrier heights: hydrogen transfer, heavy-atom transfer, nucleophilic substitution, and the unimolecular/association group. Among local functionals, only MN12-L has an MUE smaller than 4 kcal/mol on all four of these databases, and among the others MOHLYP2 and M11-L come the closest to meeting the challenge, but three global hybrid GGAs (MPW1K, B97-3, and SOGGA11-X), three long-range-corrected hybrid GGAs (LC-ωPBE, ωB97, and ωB97X), and one screened exchange hybrid GGA (N12-SX) get all four barrier height groups below 4 kcal/mol; two of these (MPW1K, LC-ωPBE) have all four below 3 kcal/mol but not below 2 kcal/mol, and three of them (B97-3, SOGGA11-X, and N12-SX) have all four below 2 kcal/mol. However, the hybrid meta-GGAs excel here; four of them (MPW1B95, M05,

M06) have a maximum of the four MUEs between 3 and 4 kcal/mol, two (MPWKCIS1K and M05-2X) have a maximum between 2 and 3 kcal/mol, and nine of them (BB1K, MPWB1K, BMK, PWB6K, M06-2X, M08-HX, M08-SO, M11, and MN12-SX) have a maximum of these four MUEs below 1.8 kcal/mol. Averaged over all 76 barrier heights, the following functionals do best: M08-HX with an MUE of 1.0 kcal/mol, M08-SO with 1.1, M06-2X, BMK, and MN12-SX with 1.2, BB1K, PWB6K and M11 with 1.3, MPWB1K with 1.4, and MPW1K and SOGGA11-X with 1.5.

### 5(f) Overall analysis

Table 15 gives the mean unsigned errors for the CE345, CS20, PE39, and PS47 databases, but only for those functionals for which we ran all four databases (which means that any functional that has a nonzero percentage of HF exchange at long range is not included). Functionals are ordered according to increasing MUE for our largest database, CE345. We highlighted in bold the performances that are better than average (for all functionals in this table) for a given database. The Pearson correlation coefficients between the columns of Table 15 are shown in Table 16. The CS20 column of Table 15 shows the best correlation with other rows, having a Pearson correlation coefficient $r$ of 0.31, 0.32, and 0.58 with the CE345, PE39, and PS47 columns, respectively. The CE345 column is the least correlated with the others, and actually has a negative correlation coefficient of $r$ = -0.10 with the PS47 column. This illustrates dramatically the difficulty of finding a functional that does well for both chemistry energetics and solid-state physics structural data.

Only seven functionals (MN12-SX, MN12-L, N12-SX, M06-L, N12, HSE06, and TPSS) have four bold entries in Table 15. N12 is the only one of these that is restricted to just the density and density gradient; N12-SX and HSE06 also have short-range HF exchange, MN12-L, M06-L and TPSS also involve kinetic energy density, while MN12-SX has both short-range HF exchange and kinetic energy density.

Table 15: Summary of the results for our four comprehensive databases (MUEs, CE345 in kcal/mol, CS20 and PS47 in Å, PE39 in eV), and average MUEs over all functionals (last row).

|  | CE345 | CS20 | PE39 | PS47 |
|---|---|---|---|---|
| **MN12-SX** | **2.16** | **0.007** | **0.29** | **0.042** |
| **MN12-L** | **2.61** | **0.008** | **0.69** | **0.034** |
| **M06-L** | **3.16** | **0.004** | **0.62** | **0.078** |
| **N12-SX** | **3.20** | **0.008** | **0.23** | **0.031** |
| **M11-L** | **3.32** | 0.010 | **0.48** | **0.066** |
| OLYP | **4.31** | **0.007** | **0.79** | 0.145 |
| **N12** | **4.40** | **0.008** | 0.81 | **0.032** |
| SOGGA11 | **4.47** | **0.006** | **0.72** | 0.116 |
| revPBE | **4.66** | 0.012 | 0.91 | 0.132 |
| RPBE | **4.71** | 0.013 | 0.91 | 0.146 |
| τ-HCTH | **4.83** | **0.005** | **0.77** | 0.114 |
| HCTH407 | **4.84** | **0.004** | **0.76** | 0.153 |
| **HSE06** | **4.88** | **0.006** | **0.23** | **0.040** |
| BPW91 | **4.88** | **0.009** | 0.92 | **0.095** |
| **TPSS** | **4.97** | **0.007** | **0.72** | **0.065** |
| revTPSS | **5.06** | **0.008** | 0.82 | **0.044** |
| BLYP | **5.13** | 0.013 | 0.98 | 0.158 |
| mPWPW | **5.17** | **0.009** | 0.90 | **0.092** |
| B86PW91 | **5.40** | 0.010 | 0.91 | 0.109 |
| PW91 | **5.40** | **0.008** | 0.90 | **0.083** |
| PBE1W | **5.43** | 0.010 | 1.00 | 0.125 |
| B86LYP | **5.76** | **0.009** | 0.97 | 0.165 |
| VSXC | **5.92** | **0.003** | 0.93 | **0.047** |
| BP86 | **6.31** | 0.011 | 0.91 | **0.096** |
| B86P86 | **6.63** | **0.009** | 0.94 | 0.104 |
| MPWLYP1W | **6.71** | 0.013 | 0.99 | 0.154 |
| PBE | **7.18** | 0.010 | **0.80** | **0.078** |
| PBELYP1W | 8.33 | 0.013 | 0.97 | 0.153 |
| TPSSLYP1W | 9.44 | 0.011 | **0.78** | 0.140 |
| MOHLYP2 | 13.81 | 0.019 | **0.77** | 0.240 |
| MOHLYP | 17.42 | 0.019 | 0.87 | 0.182 |
| PBEsol | 19.54 | 0.010 | 0.97 | **0.029** |
| SOGGA | 21.32 | 0.010 | 0.96 | **0.022** |
| SVWN3 | 30.22 | 0.010 | 1.05 | **0.053** |
| SVWN5 | 33.80 | 0.010 | 1.05 | **0.053** |
|  |  |  |  |  |
| Average | 8.15 | 0.009 | 0.81 | 0.098 |

Table 16: Pearson correlation coefficients of the columns of Table 15

|        | CE345 | CS20 | PE39 | PS47  |
|--------|-------|------|------|-------|
| CE345  | 1.00  | 0.31 | 0.43 | -0.10 |
| CS20   | 0.31  | 1.00 | 0.32 | 0.58  |
| PE39   | 0.43  | 0.32 | 1.00 | 0.31  |
| PS47   | -0.10 | 0.58 | 0.31 | 1.00  |

For the next analysis, we sorted the various density functionals by decades and calculated the average MUE on the broad chemistry CE345 database for each decade. These results are reported in Figure 4. We can clearly see significant improvement in each decade, and although there are only a few functionals from 2010 and later, they have an average MUE that is well below 3 kcal/mol and that is 3.7 times smaller than the average MUE from the 1980s functionals.

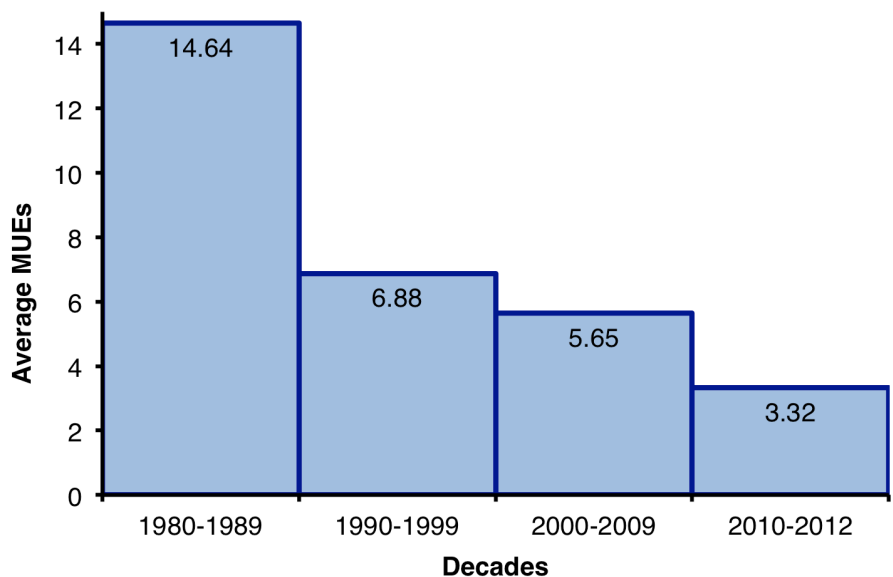

Figure 4. Average of the mean unsigned errors (MUEs, kcal/mol) for CE345 for the functionals considered in this article, sorted according to the decade in which they were published.

A completely universal functional is extremely hard to determine since it is very complicated and nonlocal, and therefore probably it will never be found [193]. Systematic approaches, such as a truncated Taylor expansion, which have guided many efforts in the past, are problematic. Quantum Monte Carlo investigations have been used to try to identify design elements for density functional theory [194,195], but so far this has not led to improved functionals. Nevertheless continued investigations of what works and what doesn't and what ranges of the variables are important for various kinds of properties [196,197] have contributed to progress, despite there being no *a priori* guarantees and no systematic path. This progress gives hope that continued exploration of functions of local variables and functionals of nonlocal functions may lead to further success.

## 6. Future developments

Where do we need to improve DFT?

(1) We need to keep improving and expanding the databases, for example on the solid-state physics side, where many important properties, such as surface energies and chemisorption and physisorption energies can be added if suitable reference data can be identified. The performances of DFT for such properties are either largely unknown, or have been investigated to a much lesser extent than the key quantities in small-molecule chemistry. The addition of such databases to the evaluation of DFT functionals can be a starting point to improve them, in the same way that the recent addition of solid-state data to the development of our own Minnesota functionals expanded their reliability and brought us a new generation of improved functionals.

(2) According to the results presented in the previous section, one property that will need major attention in the future is the treatment of multireference systems and this is confirmed by the increase of computational studies on transition metal chemistry [115]. Kohn-Sham DFT with the unknown exact functional is exact even for multireference systems, but since the density is represented by the density of a single Slater determinant, it will require greater insight and more complicated functional forms to represent the exchange-correlation energy if the orbital product in the determinant is representing the density mathematically but not physically. The best method (M11-L) for our multireference database still has a mean error of 6 kcal/mol, which is too large and will need to be improved in the future. However, the sources of these large errors are not fully understood, and more investigations will be necessary before the emergence of reliable strategies to systematically improve the treatment of multireference systems. One start on understanding multireference systems is more systematic exploration of open-shell systems [198].

## 7. Perspective

This brings us to the next question: are there prospects for further improvement? It is always hard to make predictions, "especially about the future", however we believe that there will indeed be further improvement. This may involve breaking of some of the constraints that we are following right now or—less likely as it seems to us—adding new constraints. Even more important may be rethinking some of the functional dependencies of density functionals currently available. As just one example of how to proceed, it is worth re-examining whether the UEG is the best starting point, and we believe that the next generation of functionals should look more widely for productive and more realistic models (e.g., as in the Becke–Roussel functional [199], where the H atom is used).

As far as the older Minnesota functionals, M05 and M06, the suggestions that we gave in Ref. [200] remains valid, and we report them once again here: "We recommend (1) the M06-2X, BMK, and M05-2X functionals for main-group thermochemistry and kinetics; (2) M06-2X, M05-2X, and M06 for systems where main-group thermochemistry, kinetics, and noncovalent interactions are all important; (3) the M06-L and M06 functionals for transition metal thermochemistry; (4) M06 for problems involving rearrangements of both organic and transition metal bonds; (5) M06-2X, M05-2X, M06-HF, M06, and M06-L for the study of noncovalent interactions; (6) M06-HF when the use of full HF exchange is important, for example, to avoid the error of self-interaction at long range; and (7) M06-L when a local functional is required, because a local functional has much lower cost for large systems." However, in view of the results presented in the present review, the new M11 family of Minnesota functionals and the N12 functional provide top-level performance that in most respects exceeds that of the previous Minnesota functionals. For this reason we can add M11 to the suggestion for main-group thermochemistry, kinetics, and noncovalent interactions and for problems where the use of full HF exchange at long range is important. M11-L is particularly successful for transition metal thermochemistry, kinetics, and noncovalent interactions. Functionals under development that extend what we have learned in the development of the N12 functional should be even better.

The SOGGA, SOGGA11, and SOGGA11-X functionals have been particularly important in exploring the limit of what a GGA functional form can do. However, they are usually less successful than the corresponding meta-GGA (Minnesota) functionals. Nevertheless they provided insight on which we could build, and SOGGA (suggested for solid-state physics) and SOGGA11 (suggested for chemistry) can be joined by the N12 functional, which is capable of providing balanced performance for both chemistry and solid-state physics, using the same ingredients as its predecessors and at essentially the same computational cost. Combining the nonseparable form with kinetic energy density yields the very successful local functional, MN12-L, and adding screened exchange gives the even better, MN12-SX; these functionals have outstanding overall performance in our tests so far.

## 8. Conclusion

One often hears that wave function theory is systematically improvable, while density functional theory is not. However, the facts show that since the 1980s, the mean unsigned error of density functional theory on a broad chemistry energetic database has been improved by a factor of 3.7. Furthermore we are recently beginning to see similar improvements on solid-state physics databases. Density functional theory has become the method of choice for most physical properties and chemical and materials calculations, but there are still areas where we need significant improvements, and we expect continued improvement in the years ahead.

## Appendix: Formula for the spin-unpolarized exchange density

The formulas for the exchange density functionals in the main text are those for the more general spin-polarized case. In many cases, it can be convenient to present the formula for the exchange in the spin-unpolarized case ($\rho_\alpha = \rho_\beta = \frac{1}{2}\rho$). The derivation of the spin-polarized formulas for exchange from the spin-unpolarized ones is straightforward and uses the spin-scaling relationship:

$$E_x\left[\rho_\alpha, \rho_\beta\right] = \frac{1}{2}\left\{E_x\left[2\rho_\alpha\right] + E_x\left[2\rho_\beta\right]\right\}. \tag{A1}$$

However, some of the local quantities, in particular the reduced density gradients and the UEG exchange energies, have different numerical prefactors in these cases, and the following definitions should be kept in mind when reading the literature:

<div style="text-align:center">spin-polarized:       spin-unpolarized:</div>

$$s_\sigma \equiv \frac{1}{2\left(6\pi^2\right)^{1/3}} \frac{|\nabla \rho_\sigma|}{\rho_\sigma^{4/3}} \qquad s \equiv \frac{1}{2\left(3\pi^2\right)^{1/3}} \frac{|\nabla \rho|}{\rho^{4/3}} \tag{A2}$$

$$\varepsilon_{x,\sigma}^{\mathrm{LSDA}} \equiv -\frac{3}{2}\left(\frac{3}{4\pi}\right)^{1/3} \rho_\sigma^{1/3} \qquad \varepsilon_x^{\mathrm{LDA}} \equiv -\frac{3}{4}\left(\frac{3}{\pi}\right)^{1/3} \rho^{1/3} \tag{A3}$$

$$E_x = \sum_\sigma \int dr\, \rho_\sigma\, \varepsilon_{x,\sigma}^{\mathrm{LSDA}} \qquad E_x = \int dr\, \rho\, \varepsilon_x^{\mathrm{LDA}} \tag{A4}$$

## Acknowledgments

The authors are grateful to Yan Zhao and Nate Schultz for their contributions to the Minnesota density functional program and to the many workers in density functional theory who paved the way for our own work and from whose personal comments and writings we learned many useful things, with special shout-outs to Axel Becke, Nick Handy, Mel Levy, and John Perdew. This work was supported in part by the Air Force Office of Scientific Research under grant no. FA9550-11-0078 and the National Science Foundation under grant no. CHE09-56776.

ELECTRONIC SUPPLEMENTARY MATERIAL

**The quest for a universal density functional: The accuracy of density functionals across a broad spectrum of databases in chemistry and physics**

Roberto Peverati and Donald G. Truhlar

Date of final revision: July 25, 2013

Mean unsigned errors (MUEs) results for the secondary and analytical databases are presented in this section. We grouped the functionals in the same eight classes used for the CE345 database: LSDA and first generation GGA functionals (Table S1); second generation GGA and NGA functionals (Table S2); first generation global hybrid GGA functionals (Table S3); second generation global hybrid GGA functionals (Table S4); range-separated hybrid GGA functionals (Table S5); meta-GGA functionals (Table S6); first generation hybrid meta-GGA functionals (Table S7); global and range-separated hybrid meta-GGA functionals published after M05 (Table S8). Within each table functionals are ordered according to the year of publication.

Table S1: Mean unsigned errors (MUEs, DG6 and SSLC18 in Å, all others in kcal/mol) for the secondary and analytical databases for LSDA and GGA functionals published before 1993 (year of publication of B3LYP).

| | GKSVWN5 | GKSVWN3 | B86P86 | B86LYP | BP86 | BLYP | B86PW91 | BPW91 | PW91 |
|---|---|---|---|---|---|---|---|---|---|
| Secondary Subsets: | | | | | | | | | |
| AE6/11 | 16.21 | 17.86 | 3.24 | 1.41 | 3.30 | 1.46 | 0.78 | 1.44 | 3.28 |
| SB1AE97 | 16.12 | 17.67 | 3.47 | 1.38 | 3.40 | 1.38 | 1.03 | 1.20 | 2.94 |
| LB1AE12 | 39.13 | 43.17 | 5.52 | 5.96 | 12.08 | 6.91 | 5.70 | 8.13 | 13.05 |
| IP13/03 | 5.49 | 15.85 | 6.04 | 3.56 | 5.43 | 4.81 | 4.53 | 3.74 | 4.54 |
| IPM8 | 17.03 | 27.43 | 14.12 | 9.73 | 13.94 | 9.41 | 10.45 | 10.27 | 12.03 |
| ABDE4/05 | 15.84 | 17.98 | 6.03 | 10.24 | 6.16 | 10.60 | 8.44 | 8.37 | 3.82 |
| ABDEL8 | 11.74 | 13.60 | 7.65 | 11.54 | 8.09 | 12.18 | 10.61 | 10.85 | 6.80 |
| πIE3/06 | 9.30 | 9.32 | 9.05 | 8.97 | 8.78 | 8.68 | 8.94 | 8.68 | 8.83 |
| PA-CP5/06 | 3.88 | 3.31 | 9.45 | 9.62 | 4.92 | 5.67 | 11.15 | 6.63 | 4.48 |
| PA-SB5/06 | 3.03 | 3.21 | 12.05 | 11.62 | 5.04 | 4.91 | 13.77 | 6.57 | 5.12 |
| HATBH12/08 | 23.08 | 22.84 | 12.40 | 11.74 | 15.18 | 14.33 | 9.88 | 12.66 | 14.97 |
| NSBH16/08 | 8.52 | 8.65 | 6.87 | 7.43 | 6.94 | 7.37 | 5.96 | 6.04 | 7.59 |
| UABH10/08 | 5.87 | 5.73 | 4.40 | 3.77 | 3.79 | 3.43 | 3.44 | 2.74 | 3.33 |
| HB6/04 | 4.65 | 4.89 | 0.95 | 0.75 | 0.72 | 1.18 | 0.54 | 1.64 | 0.70 |
| CT7/04 | 6.79 | 7.00 | 3.11 | 2.83 | 2.04 | 1.67 | 2.26 | 1.39 | 3.35 |
| DI6/04 | 2.93 | 3.08 | 0.52 | 0.49 | 0.66 | 1.00 | 0.69 | 1.17 | 0.62 |
| EDCE19 | 4.90 | 5.10 | 1.61 | 1.44 | 1.19 | 1.30 | 1.22 | 1.40 | 1.65 |
| W17/05 | 0.30 | 0.33 | 0.36 | 0.23 | 0.65 | 0.45 | 0.41 | 0.71 | 0.27 |
| PPS5/05 | 0.60 | 0.71 | 1.71 | 2.05 | 3.63 | 4.01 | 2.25 | 4.17 | 1.81 |
| Analytical Subsets: | | | | | | | | | |
| MBE18 | 23.74 | 26.96 | 9.36 | 7.47 | 8.11 | 7.10 | 10.62 | 7.55 | 9.58 |
| TMBE15 | 26.31 | 29.69 | 10.82 | 8.66 | 9.01 | 7.39 | 12.04 | 8.29 | 10.85 |
| DBH24/08 | 13.39 | 13.32 | 7.75 | 6.99 | 8.43 | 7.67 | 6.26 | 6.92 | 8.45 |
| DG6 | 0.012 | 0.011 | 0.006 | 0.007 | 0.015 | 0.022 | 0.009 | 0.013 | 0.012 |
| SSLC18 | 0.056 | 0.056 | 0.084 | 0.122 | 0.071 | 0.113 | 0.092 | 0.076 | 0.081 |

Table S2: Mean unsigned errors (MUEs, DG6 and SSLC18 in Å, all others in kcal/mol) for the secondary and analytical databases for GGA and NGA functionals published after 1993 (year of publication of B3LYP).

| | PBE | HCTH407 | mPWPW | revPBE | RPBE | OLYP | MPWLYP1W | PBE1W | PBELYP1W | MOHLYP | B97-D | SOGGA | PBEsol | MOHLYP2 | SOGGA11 | N12 |
|---|---|---|---|---|---|---|---|---|---|---|---|---|---|---|---|---|
| Secondary Subsets: | | | | | | | | | | | | | | | | |
| AE6/11 | 3.16 | 1.15 | 2.12 | 1.87 | 2.08 | 0.86 | 1.24 | 2.39 | 1.11 | 2.33 | 0.38 | 7.53 | 7.45 | 18.18 | 1.78 | 0.91 |
| SB1AE97 | 2.77 | 1.01 | 1.80 | 1.57 | 1.91 | 0.79 | 1.12 | 2.20 | 1.49 | 2.48 | 0.76 | 7.36 | 7.51 | 17.63 | 1.47 | 1.12 |
| LB1AE12 | 12.98 | 4.71 | 10.32 | 5.36 | 4.83 | 4.64 | 8.32 | 10.86 | 8.40 | 3.87 | 3.64 | 23.27 | 22.67 | 28.17 | 8.38 | 6.03 |
| IP13/03 | 3.62 | 5.46 | 4.19 | 3.07 | 3.12 | 2.60 | 4.79 | 5.37 | 6.19 | 4.25 | 2.70 | 2.48 | 2.66 | 11.59 | 4.74 | 3.14 |
| IPM8 | 10.41 | 9.00 | 11.22 | 7.98 | 7.80 | 3.20 | 11.49 | 12.57 | 14.16 | 2.96 | 4.38 | 8.30 | 10.79 | 7.26 | 8.54 | 3.95 |
| ABDE4/05 | 4.09 | 9.05 | 6.31 | 10.59 | 11.21 | 10.28 | 9.35 | 6.45 | 9.62 | 14.09 | 7.67 | 5.09 | 4.03 | 32.29 | 5.00 | 3.81 |
| ABDEL8 | 7.16 | 11.61 | 9.01 | 12.94 | 13.49 | 12.12 | 10.95 | 8.84 | 10.92 | 15.24 | 9.13 | 3.87 | 3.18 | 32.92 | 7.89 | 6.54 |
| πIE3/06 | 8.78 | 8.43 | 8.76 | 8.58 | 8.56 | 8.69 | 8.72 | 8.68 | 8.60 | 8.41 | 8.36 | 9.05 | 9.03 | 8.04 | 8.86 | 9.14 |
| PA-CP5/06 | 4.34 | 7.94 | 5.65 | 7.15 | 7.34 | 8.45 | 5.50 | 5.39 | 6.03 | 7.02 | 9.32 | 1.64 | 1.91 | 13.58 | 5.23 | 7.63 |
| PA-SB5/06 | 4.92 | 8.38 | 5.76 | 6.30 | 6.24 | 7.82 | 4.76 | 5.38 | 5.13 | 5.27 | 8.59 | 3.47 | 3.59 | 9.72 | 8.73 | 9.28 |
| HATBH12/08 | 14.60 | 8.52 | 13.77 | 11.89 | 11.77 | 10.91 | 15.00 | 14.07 | 14.64 | 11.35 | 9.39 | 17.66 | 17.82 | 3.60 | 7.80 | 12.02 |
| NSBH16/08 | 7.01 | 7.22 | 6.85 | 5.64 | 5.74 | 2.77 | 7.97 | 6.91 | 7.64 | 3.28 | 6.10 | 7.15 | 7.40 | 3.51 | 3.18 | 5.66 |
| UABH10/08 | 3.27 | 2.14 | 3.02 | 2.60 | 2.59 | 2.45 | 3.62 | 3.23 | 3.61 | 3.19 | 3.04 | 4.13 | 4.24 | 3.01 | 1.98 | 2.59 |
| HB6/04 | 0.45 | 1.69 | 0.57 | 2.01 | 1.90 | 3.60 | 0.73 | 0.55 | 0.82 | 4.50 | 0.40 | 1.78 | 1.69 | 8.78 | 3.23 | 0.49 |
| CT7/04 | 2.97 | 1.31 | 2.26 | 1.28 | 1.35 | 1.55 | 2.45 | 2.51 | 2.44 | 1.80 | 1.89 | 4.29 | 4.21 | 5.29 | 1.35 | 1.87 |
| DI6/04 | 0.45 | 0.55 | 0.55 | 1.09 | 0.86 | 2.35 | 0.42 | 0.40 | 0.33 | 2.76 | 0.35 | 1.27 | 1.16 | 5.39 | 0.45 | 0.85 |
| EDCE19 | 1.38 | 1.19 | 1.19 | 1.45 | 1.37 | 2.45 | 1.27 | 1.23 | 1.26 | 2.95 | 0.93 | 2.55 | 2.45 | 6.42 | 1.66 | 1.11 |
| WI7/05 | 0.14 | 0.29 | 0.24 | 0.32 | 0.24 | 0.39 | 0.23 | 0.16 | 0.21 | 0.47 | 0.15 | 0.13 | 0.10 | 0.81 | 0.91 | 0.42 |
| PPS5/05 | 2.23 | 2.14 | 2.99 | 3.91 | 3.54 | 4.97 | 2.99 | 2.49 | 2.58 | 6.05 | 0.23 | 1.55 | 1.66 | 8.46 | 0.36 | 3.25 |
| Analytical Subsets: | | | | | | | | | | | | | | | | |
| MBE18 | 7.88 | 7.83 | 7.26 | 6.42 | 6.33 | 8.78 | 8.54 | 8.80 | 7.95 | 9.29 | 26.96 | 12.20 | 12.15 | 25.31 | 11.83 | 8.62 |
| TMBE15 | 8.82 | 8.85 | 8.02 | 6.60 | 6.45 | 9.41 | 9.27 | 9.92 | 8.50 | 9.51 | 29.69 | 13.82 | 13.72 | 25.71 | 13.32 | 8.83 |
| DBH24/08 | 8.18 | 7.01 | 7.66 | 6.39 | 6.32 | 5.34 | 8.13 | 7.74 | 7.85 | 5.67 | 5.92 | 10.22 | 10.25 | 3.05 | 5.14 | 6.43 |
| DG6 | 0.013 | 0.004 | 0.013 | 0.016 | 0.017 | 0.009 | 0.021 | 0.015 | 0.021 | 0.024 | 0.013 | 0.009 | 0.010 | 0.025 | 0.008 | 0.008 |
| SSLC18 | 0.067 | 0.12 | 0.07 | 0.108 | 0.109 | 0.121 | 0.116 | 0.099 | 0.108 | 0.134 | - | 0.021 | 0.025 | 0.076 | 0.122 | 0.021 |

Table S3: Mean unsigned errors (MUEs, DG6 and SSLC18 in Å, all others in kcal/mol) for the secondary and analytical databases for global hybrid GGA functionals published before 1993 (year of publication of B3LYP).

| | HF | MP2 | HFLYP | HFPW91 | B3PW91 | B3LYP |
|---|---|---|---|---|---|---|
| Secondary Subsets: | | | | | | |
| AE6/11 | 31.25 | 2.09 | 7.93 | 12.34 | 0.70 | 0.75 |
| SB1AE97 | 29.57 | 1.99 | 7.03 | 10.59 | 0.67 | 0.95 |
| LB1AE12 | 77.03 | 4.72 | 41.13 | 55.39 | 2.55 | 2.15 |
| IP13/03 | 17.88 | 3.52 | 5.05 | 21.01 | 4.30 | 4.76 |
| IPM8 | 22.65 | 19.72 | 16.05 | 34.77 | 5.79 | 7.27 |
| ABDE4/05 | 35.42 | 5.13 | 11.35 | 24.82 | 7.43 | 8.73 |
| ABDEL8 | 36.68 | 4.64 | 12.62 | 23.05 | 9.79 | 10.40 |
| πIE3/06 | 2.67 | 6.39 | 2.48 | 2.86 | 6.18 | 6.25 |
| PA-CP5/06 | 12.83 | 1.27 | 11.50 | 22.22 | 6.97 | 5.93 |
| PA-SB5/06 | 11.01 | 0.73 | 12.10 | 19.59 | 7.52 | 6.01 |
| HATBH12/08 | 17.19 | 12.07 | 12.27 | 18.56 | 6.86 | 8.16 |
| NSBH16/08 | 6.63 | 0.76 | 5.29 | 5.55 | 2.27 | 3.43 |
| UABH10/08 | 3.88 | 5.50 | 3.57 | 5.07 | 1.90 | 2.02 |
| HB6/04 | 2.23 | 0.26 | 1.50 | 2.24 | 1.03 | 0.60 |
| CT7/04 | 3.77 | 0.73 | 0.87 | 3.84 | 0.63 | 0.71 |
| DI6/04 | 2.39 | 0.45 | 0.22 | 2.61 | 0.97 | 0.78 |
| EDCE19 | 2.85 | 0.50 | 0.87 | 2.95 | 0.86 | 0.70 |
| WI7/05 | 0.31 | 0.07 | 0.15 | 0.26 | 0.52 | 0.30 |
| PPS5/05 | 2.76 | 1.27 | 0.47 | 1.67 | 3.11 | 2.89 |
| Analytical Subsets: | | | | | | |
| MBE18 | 71.67 | 56.41 | 49.90 | 60.51 | 14.54 | 13.14 |
| TMBE15 | 79.02 | 69.47 | 57.71 | 70.27 | 16.53 | 14.94 |
| DBH24/08 | 10.18 | 5.63 | 6.69 | 10.74 | 3.74 | 4.18 |
| DG6 | 0.015 | 0.007 | 0.031 | 0.036 | 0.011 | 0.009 |
| SSLC18 | - | - | - | - | - | - |

Table S4: Mean unsigned errors (MUEs, DG6 and SSLC18 in Å, all others in kcal/mol) for the secondary and analytical databases for global hybrid GGA functionals published after 1993 (year of publication of B3LYP).

| | PBE0 | mPW1PW | B1LYP | B98 | MPW1K | O3LYP | MPW3LYP | MPWLYP1M | B97-3 | SOGGA11-X |
|---|---|---|---|---|---|---|---|---|---|---|
| Secondary Subsets: | | | | | | | | | | |
| AE6/11 | 1.20 | 1.02 | 2.70 | 0.69 | 2.47 | 0.52 | 0.31 | 1.02 | 0.63 | 0.72 |
| SB1AE97 | 0.92 | 0.94 | 2.67 | 0.67 | 2.11 | 0.75 | 0.65 | 0.99 | 0.62 | 0.65 |
| LB1AE12 | 2.80 | 3.75 | 6.51 | 2.22 | 13.04 | 2.97 | 2.14 | 6.26 | 1.98 | 3.28 |
| IP13/03 | 3.24 | 3.76 | 3.75 | 3.23 | 3.58 | 2.47 | 4.35 | 4.51 | 3.54 | 3.20 |
| IPM8 | 3.57 | 4.22 | 2.71 | 4.10 | 3.47 | 4.87 | 6.24 | 8.98 | 3.36 | 3.96 |
| ABDE4/05 | 5.14 | 6.94 | 10.57 | 4.99 | 7.55 | 9.13 | 7.25 | 8.65 | 4.74 | 4.68 |
| ABDEL8 | 8.11 | 9.59 | 12.16 | 7.71 | 10.12 | 10.93 | 9.12 | 10.44 | 7.70 | 5.12 |
| πIE3/06 | 5.65 | 5.64 | 5.66 | 6.26 | 3.58 | 7.19 | 6.12 | 8.13 | 5.39 | 3.86 |
| PA-CP5/06 | 5.86 | 6.90 | 6.22 | 7.38 | 7.95 | 8.31 | 5.20 | 5.12 | 7.81 | 7.28 |
| PA-SB5/06 | 6.64 | 7.52 | 6.08 | 7.39 | 8.83 | 8.18 | 5.31 | 4.55 | 8.11 | 6.21 |
| HATBH12/08 | 6.30 | 5.66 | 6.45 | 4.85 | 1.75 | 7.94 | 8.56 | 13.73 | 2.19 | 1.86 |
| NSBH16/08 | 2.07 | 8.07 | 2.69 | 3.00 | 1.26 | 1.31 | 3.77 | 7.11 | 0.75 | 0.54 |
| UABH10/08 | 2.15 | 1.99 | 1.75 | 1.97 | 2.41 | 2.22 | 2.16 | 3.23 | 1.42 | 1.33 |
| HB6/04 | 0.40 | 0.39 | 0.72 | 0.45 | 0.33 | 2.77 | 0.56 | 0.62 | 1.16 | 0.24 |
| CT7/04 | 1.06 | 0.66 | 0.49 | 0.92 | 0.44 | 1.18 | 1.38 | 2.29 | 0.49 | 0.21 |
| DI6/04 | 0.35 | 0.52 | 0.93 | 0.34 | 0.52 | 1.97 | 0.29 | 0.36 | 0.82 | 0.54 |
| EDCE19 | 0.63 | 0.53 | 0.70 | 0.59 | 0.43 | 1.93 | 0.78 | 1.16 | 0.81 | 0.32 |
| WI7/05 | 0.13 | 0.22 | 0.30 | 0.13 | 0.20 | 0.33 | 0.20 | 0.23 | 0.48 | 0.42 |
| PPS5/05 | 1.70 | 2.23 | 2.93 | 1.96 | 1.70 | 4.15 | 1.95 | 2.68 | 2.33 | 2.07 |
| Analytical Subsets: | | | | | | | | | | |
| MBE18 | 16.37 | 17.29 | 16.69 | 10.03 | 31.14 | 11.22 | 15.06 | 6.19 | 13.15 | 20.55 |
| TMBE15 | 18.72 | 19.74 | 18.99 | 11.64 | 36.73 | 12.37 | 17.67 | 6.47 | 15.43 | 23.76 |
| DBH24/08 | 3.58 | 7.32 | 3.23 | 3.41 | 1.48 | 3.87 | 4.49 | 7.39 | 1.83 | 1.29 |
| DG6 | 0.003 | 0.003 | 0.009 | 0.007 | 0.011 | 0.004 | 0.009 | 0.018 | 0.004 | 0.005 |
| SSLC18 | - | - | - | - | - | - | - | - | - | - |

Table S5: Mean unsigned errors (MUEs, DG6 and SSLC18 in Å, all others in kcal/mol) for the secondary and analytical databases for range-separated hybrid GGA functionals.

| | CAM-B3LYP | LC-ωPBE | HSE06 | ωB97 | ωB97X | ωB97X-D | N12-SX |
|---|---|---|---|---|---|---|---|
| Secondary Subsets: | | | | | | | |
| AE6/11 | 0.36 | 1.04 | 0.94 | 0.57 | 0.37 | 0.39 | 0.79 |
| SB1AE97 | 0.65 | 0.82 | 0.82 | 0.52 | 0.45 | 0.47 | 0.68 |
| LB1AE12 | 2.40 | 3.28 | 2.81 | 2.94 | 2.45 | 2.24 | 3.43 |
| IP13/03 | 4.70 | 5.45 | 3.23 | 2.41 | 2.82 | 3.02 | 3.06 |
| IPM8 | 5.72 | 7.09 | 5.27 | 9.93 | 6.55 | 3.39 | 5.52 |
| ABDE4/05 | 5.55 | 4.87 | 5.82 | 3.76 | 3.70 | 3.73 | 3.73 |
| ABDEL8 | 7.42 | 7.37 | 8.70 | 3.89 | 4.82 | 4.92 | 6.08 |
| πIE3/06 | 3.15 | 0.79 | 5.91 | 0.84 | 1.16 | 3.07 | 6.03 |
| PA-CP5/06 | 3.65 | 4.80 | 5.89 | 5.33 | 5.40 | 7.03 | 7.46 |
| PA-SB5/06 | 4.05 | 5.84 | 6.70 | 4.39 | 5.25 | 7.35 | 9.44 |
| HATBH12/08 | 5.24 | 2.02 | 6.70 | 2.79 | 2.06 | 1.90 | 3.65 |
| NSBH16/08 | 1.20 | 2.64 | 2.55 | 1.68 | 3.84 | 6.27 | 2.21 |
| UABH10/08 | 1.73 | 2.32 | 2.05 | 3.10 | 2.36 | 1.90 | 2.83 |
| HB6/04 | 0.56 | 0.58 | 0.48 | 1.02 | 0.94 | 0.44 | 0.60 |
| CT7/04 | 0.48 | 0.90 | 1.31 | 0.56 | 0.67 | 0.28 | 0.96 |
| DI6/04 | 0.40 | 0.78 | 0.34 | 0.64 | 0.58 | 0.27 | 0.36 |
| EDCE19 | 0.48 | 0.76 | 0.74 | 0.73 | 0.73 | 0.33 | 0.65 |
| WI7/05 | 0.14 | 0.23 | 0.13 | 0.07 | 0.03 | 0.04 | 0.25 |
| PPS5/05 | 1.87 | 1.66 | 1.65 | 0.35 | 0.29 | 0.69 | 1.75 |
| Analytical Subsets: | | | | | | | |
| MBE18 | 16.88 | 21.77 | 15.26 | 13.44 | 14.76 | 14.00 | 6.74 |
| TMBE15 | 19.45 | 25.26 | 17.40 | 15.67 | 17.35 | 16.15 | 7.40 |
| DBH24/08 | 2.56 | 2.12 | 3.77 | 2.63 | 3.51 | 4.21 | 3.14 |
| DG6 | 0.007 | 0.012 | 0.003 | 0.007 | 0.005 | 0.003 | 0.005 |
| SSLC18 | - | - | 0.036 | - | - | - | 0.022 |

Table S6: Mean unsigned errors (MUEs, DG6 and SSLC18 in Å, all others in kcal/mol) for the secondary and analytical databases for meta-GGA functionals.

| | VSXC | τ-HCTH | TPSS | TPSSLYP1W | M06-L | revTPSS | M11-L | MN12-L |
|---|---|---|---|---|---|---|---|---|
| Secondary Subsets: | | | | | | | | |
| AE6/11 | 0.57 | 0.80 | 1.02 | 2.47 | 0.64 | 1.17 | 1.00 | 0.65 |
| SB1AE97 | 0.67 | 0.79 | 0.99 | 2.52 | 0.83 | 0.85 | 0.68 | 0.63 |
| LB1AE12 | 2.19 | 4.52 | 3.70 | 3.64 | 2.25 | 4.13 | 2.94 | 2.72 |
| IP13/03 | 3.34 | 4.40 | 3.09 | 4.53 | 3.08 | 2.96 | 3.11 | 2.71 |
| IPM8 | 4.53 | 4.81 | 5.84 | 8.28 | 5.21 | 5.47 | 7.03 | 4.79 |
| ABDE4/05 | 8.14 | 7.93 | 9.56 | 13.01 | 5.54 | 7.64 | 5.14 | 4.25 |
| ABDEL8 | 8.12 | 10.84 | 10.93 | 13.81 | 8.85 | 9.02 | 6.98 | 5.16 |
| πIE3/06 | 8.45 | 8.58 | 8.44 | 8.40 | 5.43 | 8.45 | 7.65 | 7.50 |
| PA-CP5/06 | 8.47 | 8.91 | 8.57 | 9.41 | 7.94 | 8.67 | 4.79 | 4.14 |
| PA-SB5/06 | 8.24 | 8.71 | 7.47 | 7.23 | 6.19 | 6.67 | 3.98 | 5.19 |
| HATBH12/08 | 7.10 | 8.87 | 14.32 | 13.11 | 5.74 | 14.50 | 4.59 | 3.89 |
| NSBH16/08 | 4.95 | 5.64 | 7.94 | 9.29 | 3.59 | 8.05 | 2.11 | 1.41 |
| UABH10/08 | 2.40 | 2.76 | 3.96 | 3.41 | 1.86 | 4.17 | 1.99 | 1.59 |
| HB6/04 | 0.45 | 0.67 | 0.45 | 0.66 | 0.22 | 0.42 | 0.78 | 0.73 |
| CT7/04 | 2.86 | 1.78 | 2.22 | 2.36 | 1.78 | 2.36 | 0.94 | 0.70 |
| DI6/04 | 1.08 | 0.50 | 0.52 | 0.39 | 0.32 | 0.46 | 0.30 | 0.42 |
| EDCE19 | 1.54 | 1.03 | 1.12 | 1.20 | 0.83 | 1.15 | 0.69 | 0.62 |
| WI7/05 | 0.92 | 0.23 | 0.20 | 0.21 | 0.17 | 0.18 | 0.31 | 0.18 |
| PPS5/05 | 6.62 | 2.63 | 2.74 | 2.73 | 0.21 | 2.45 | 0.43 | 0.27 |
| Analytical Subsets: | | | | | | | | |
| MBE18 | 8.50 | 11.28 | 6.44 | 7.12 | 8.04 | 7.24 | 6.38 | 10.30 |
| TMBE15 | 9.46 | 13.17 | 7.08 | 7.65 | 9.12 | 7.86 | 6.65 | 11.25 |
| DBH24/08 | 4.31 | 5.78 | 8.20 | 7.35 | 4.05 | 8.14 | 2.56 | 1.87 |
| DG6 | 0.006 | 0.006 | 0.011 | 0.019 | 0.006 | 0.011 | 0.011 | 0.005 |
| SSLC18 | 0.076 | 0.107 | 0.054 | 0.076 | 0.071 | 0.034 | 0.050 | 0.002 |

Table S7: Mean unsigned errors (MUEs, DG6 and SSLC18 in Å, all others in kcal/mol) for the secondary and analytical databases for hybrid meta-GGA functionals published before M05.

| | TPSSh | τ-HCTHhyb | BB1K | MPWB1K | MPW1B95 | BMK | TPSS1KCIS | MPWKCIS1K | MPW1KCIS | PBE1KCIS | PWB6K | PW6B95 |
|---|---|---|---|---|---|---|---|---|---|---|---|---|
| Secondary Subsets: | | | | | | | | | | | | |
| AE6/11 | 1.21 | 0.89 | 1.37 | 1.06 | 0.84 | 0.47 | 0.58 | 1.64 | 2.35 | 2.66 | 1.42 | 0.44 |
| SB1AE97 | 0.94 | 0.74 | 1.23 | 0.84 | 0.63 | 0.48 | 0.67 | 1.31 | 2.13 | 2.40 | 1.28 | 0.42 |
| LB1AE12 | 4.05 | 2.99 | 7.80 | 7.53 | 2.31 | 1.77 | 2.29 | 9.04 | 5.20 | 3.85 | 9.09 | 1.92 |
| IP13/03 | 3.15 | 4.09 | 2.10 | 3.09 | 1.92 | 4.22 | 2.57 | 2.57 | 3.07 | 2.81 | 2.60 | 3.45 |
| IPM8 | 3.71 | 3.95 | 3.51 | 3.60 | 1.93 | 3.87 | 3.72 | 2.98 | 6.23 | 4.08 | 3.36 | 2.88 |
| ABDE4/05 | 9.53 | 4.36 | 4.25 | 4.18 | 4.01 | 3.57 | 8.30 | 6.34 | 5.67 | 4.51 | 4.19 | 3.98 |
| ABDEL8 | 10.94 | 7.46 | 6.11 | 5.07 | 4.85 | 3.88 | 10.36 | 8.84 | 8.27 | 7.06 | 5.53 | 6.08 |
| πIE3/06 | 7.21 | 6.99 | 3.60 | 3.43 | 4.93 | 3.90 | 6.85 | 3.68 | 6.80 | 5.93 | 3.19 | 5.28 |
| PA-CP5/06 | 8.71 | 7.63 | 6.24 | 5.81 | 5.07 | 4.39 | 7.95 | 6.27 | 4.83 | 4.31 | 6.14 | 5.72 |
| PA-SB5/06 | 7.92 | 7.54 | 7.19 | 6.90 | 5.96 | 5.16 | 7.41 | 7.40 | 5.51 | 5.23 | 7.15 | 6.25 |
| HATBH12/08 | 11.18 | 6.27 | 1.57 | 1.65 | 4.29 | 1.17 | 8.94 | 2.54 | 9.45 | 7.87 | 1.75 | 5.03 |
| NSBH16/08 | 5.79 | 4.57 | 1.27 | 1.18 | 1.17 | 0.86 | 4.88 | 1.66 | 3.00 | 1.92 | 1.09 | 2.06 |
| UABH10/08 | 3.18 | 2.19 | 1.44 | 1.61 | 1.31 | 1.58 | 2.09 | 2.49 | 2.31 | 2.26 | 1.53 | 1.43 |
| HB6/04 | 0.41 | 0.30 | 0.99 | 0.41 | 0.50 | 0.69 | 0.49 | 0.50 | 0.76 | 0.45 | 0.44 | 0.53 |
| CT7/04 | 1.44 | 1.38 | 0.67 | 0.23 | 0.48 | 0.43 | 1.23 | 0.48 | 0.95 | 0.99 | 0.26 | 0.70 |
| DI6/04 | 0.49 | 0.45 | 1.03 | 0.52 | 0.50 | 0.77 | 0.46 | 0.60 | 0.54 | 0.33 | 0.24 | 0.40 |
| EDCE19 | 0.82 | 0.74 | 0.89 | 0.38 | 0.49 | 0.62 | 0.75 | 0.52 | 0.76 | 0.61 | 0.31 | 0.55 |
| WI7/05 | 0.19 | 0.17 | 0.32 | 0.08 | 0.09 | 0.80 | 0.17 | 0.18 | 0.21 | 0.13 | 0.18 | 0.14 |
| PPS5/05 | 2.48 | 2.20 | 1.42 | 0.66 | 1.06 | 2.39 | 2.33 | 1.81 | 2.60 | 1.75 | 0.23 | 0.82 |
| Analytical Subsets: | | | | | | | | | | | | |
| MBE18 | 9.69 | 7.50 | 28.43 | 27.38 | 19.08 | 18.35 | 11.62 | 28.72 | 11.38 | 15.53 | 25.95 | 15.33 |
| TMBE15 | 10.78 | 8.66 | 34.22 | 32.66 | 22.67 | 21.69 | 13.42 | 34.17 | 13.31 | 18.52 | 30.29 | 17.89 |
| DBH24/08 | 6.36 | 4.38 | 1.20 | 1.24 | 2.35 | 1.22 | 4.81 | 1.86 | 5.04 | 4.22 | 1.25 | 2.73 |
| DG6 | 0.006 | 0.006 | 0.011 | 0.012 | 0.005 | 0.007 | 0.005 | 0.009 | 0.006 | 0.003 | 0.013 | 0.004 |
| SSLC18 | - | - | - | - | - | - | - | - | - | - | - | - |

Table S8: Mean unsigned errors (MUEs, DG6 and SSLC18 in Å, all others in kcal/mol) for the secondary and analytical databases for global and range-separated hybrid meta-GGA functionals published after M05.

| | M05 | M05-2X | M06-HF | M06 | M06-2X | M08-HX | M08-SO | M11 | MN12-SX |
|---|---|---|---|---|---|---|---|---|---|
| Secondary Subsets: | | | | | | | | | |
| AE6/11 | 0.45 | 0.64 | 0.83 | 0.46 | 0.31 | 0.63 | 0.40 | 0.40 | 0.56 |
| SB1AE97 | 0.50 | 0.47 | 0.57 | 0.56 | 0.41 | 0.67 | 0.59 | 0.45 | 0.47 |
| LB1AE12 | 2.89 | 2.90 | 5.66 | 2.32 | 2.17 | 2.42 | 2.36 | 2.58 | 2.22 |
| IP13/03 | 2.82 | 3.55 | 3.80 | 3.28 | 2.56 | 3.42 | 3.58 | 3.64 | 3.17 |
| IPM8 | 8.95 | 5.86 | 13.32 | 5.31 | 3.24 | 6.05 | 3.53 | 16.40 | 9.66 |
| ABDE4/05 | 5.98 | 2.45 | 4.43 | 2.84 | 2.12 | 2.67 | 2.51 | 2.45 | 3.42 |
| ABDEL8 | 8.78 | 2.73 | 4.56 | 4.72 | 2.69 | 2.87 | 3.88 | 3.48 | 4.03 |
| πIE3/06 | 1.86 | 2.96 | 1.08 | 2.00 | 1.63 | 2.74 | 2.04 | 1.36 | 5.75 |
| PA-CP5/06 | 8.05 | 2.19 | 3.77 | 5.78 | 0.79 | 0.60 | 1.61 | 1.23 | 1.99 |
| PA-SB5/06 | 5.64 | 4.00 | 0.90 | 4.45 | 2.10 | 2.61 | 1.96 | 3.78 | 2.98 |
| HATBH12/08 | 3.64 | 2.17 | 4.56 | 3.26 | 1.31 | 1.53 | 1.55 | 1.56 | 1.93 |
| NSBH16/08 | 0.78 | 1.54 | 1.64 | 2.03 | 1.36 | 1.12 | 1.04 | 1.11 | 0.96 |
| UABH10/08 | 2.23 | 1.83 | 1.51 | 1.69 | 0.91 | 1.00 | 1.15 | 1.20 | 1.29 |
| HB6/04 | 0.57 | 0.39 | 0.63 | 0.26 | 0.43 | 0.47 | 0.46 | 0.37 | 0.44 |
| CT7/04 | 0.66 | 0.45 | 0.35 | 1.07 | 0.37 | 0.59 | 0.66 | 0.30 | 0.40 |
| DI6/04 | 0.23 | 0.23 | 0.53 | 0.27 | 0.22 | 0.25 | 0.29 | 0.33 | 0.34 |
| EDCE19 | 0.50 | 0.36 | 0.50 | 0.56 | 0.34 | 0.44 | 0.48 | 0.33 | 0.40 |
| WI7/05 | 0.14 | 0.10 | 0.22 | 0.18 | 0.18 | 0.09 | 0.09 | 0.09 | 0.13 |
| PPS5/05 | 0.98 | 0.22 | 0.38 | 0.18 | 0.21 | 0.36 | 0.30 | 0.22 | 0.19 |
| Analytical Subsets: | | | | | | | | | |
| MBE18 | 9.52 | 22.99 | 40.42 | 11.71 | 25.30 | 26.58 | 25.97 | 24.64 | 11.21 |
| TMBE15 | 10.75 | 27.12 | 47.34 | 13.62 | 29.89 | 31.29 | 30.80 | 28.99 | 12.70 |
| DBH24/08 | 2.49 | 1.80 | 2.53 | 2.39 | 0.98 | 1.12 | 1.05 | 1.26 | 1.25 |
| DG6 | 0.007 | 0.006 | 0.012 | 0.006 | 0.003 | 0.005 | 0.007 | 0.007 | 0.003 |
| SSLC18 | - | - | - | - | - | - | - | - | 0.025 |